# Near-intrinsic photo- and electroluminescence from single-walled carbon nanotube thin films on BCB-passivated surfaces

*Nicolas F. Zorn, Simon Settele, Shen Zhao, Sebastian Lindenthal, Abdurrahman Ali El Yumin, Tim Wedl, Han Li, Benjamin S. Flavel, Alexander Högele, and Jana Zaumseil**

N.F. Zorn, S. Settele, S. Lindenthal, A.A. El Yumin, J. Zaumseil
Institute for Physical Chemistry, Ruprecht-Karls-Universität Heidelberg, D-69120 Heidelberg, Germany
E-mail: zaumseil@uni-heidelberg.de

S. Zhao, T. Wedl, A. Högele
Fakultät für Physik, Munich Quantum Center and Center for NanoScience (CeNS), Ludwig-Maximilians-Universität München, D-80539 München, Germany

H. Li, B.S. Flavel
Institute of Nanotechnology, Karlsruhe Institute of Technology, D-76131 Karlsruhe, Germany

Keywords: single-walled carbon nanotubes; photoluminescence; defects; electroluminescence; light-emitting field-effect transistor

Their outstanding electrical and optical properties make semiconducting single-walled carbon nanotubes (SWCNTs) highly suitable for charge transport and emissive layers in near-infrared optoelectronic devices. However, the luminescence spectra of SWCNT thin films on commonly used glass and Si/$SiO_2$ substrates are often compromised by broadening of the main excitonic emission and unwanted low-energy sidebands. Here, we demonstrate that surface passivation with the commercially available, low dielectric constant, cross-linked bis-benzocyclobutene-based polymer BCB significantly enhances the emission properties of SWCNTs to the same level as hexagonal boron nitride (*h*-BN) flakes do. The presence of BCB suppresses sideband emission, especially from the $Y_1$ band, which we attribute to defects introduced by interaction of the nanotube lattice with oxygen-containing terminal groups of the substrate surface. The facile and reproducible deposition of homogeneous BCB films over large areas, combined with their resistance against common solvents and chemicals employed during photolithography, make them compatible with standard semiconductor device fabrication. Utilizing this approach, we fabricate light-emitting (6,5) SWCNT network field-effect transistors on BCB-treated glass substrates with excellent electrical characteristics and near-intrinsic electroluminescence. Hence, we propose passivation with BCB as a standard treatment for substrates used for spectroscopic investigations of, and optoelectronic devices with, SWCNTs and other low-dimensional emitters.

## 1. Introduction

Semiconducting single-walled carbon nanotubes (SWCNTs) are a promising material for light-emitting optoelectronic devices in the near-infrared (NIR) due to their diameter-dependent, narrow-band emission peaks and high ambipolar charge carrier mobilities.[1-2] The recent progress in the purification and sorting of SWCNT raw materials, *e.g.*, by polymer wrapping in organic solvents[3-5] or aqueous two-phase extraction,[6-7] enables the preparation of purely semiconducting or even monochiral nanotube samples that are suitable for device fabrication such as light-emitting diodes and transistors based on SWCNT thin films.[8-11]

Due to their low dimensionality and hence poor screening of Coulomb interactions, the photophysical properties of SWCNTs are governed by strongly bound excitons that exhibit binding energies on the order of a few hundred millielectronvolts.[12] These excitons are extremely sensitive to their environment, such as the choice of surfactant and solvent for dispersion, or the interaction with other nanotubes in a network or with a substrate.[13] Optical transition energies of solution-dispersed SWCNTs are typically red-shifted compared to air-suspended nanotubes of the same type.[14] Furthermore, for individual SWCNTs or nanotube thin films on substrates (for a schematic illustration of a polymer-wrapped (6,5) SWCNT, see **Figure 1a**), emission efficiencies are generally low due to inter-nanotube and nanotube-substrate interactions that facilitate exciton quenching.[5,9] This behavior is often observed for commonly used Si/$SiO_2$ or glass substrates with polar terminal groups (**Figure 1b**). For example, the initially bright emission from air-suspended SWCNTs is strongly quenched upon contact with a $SiO_2$ substrate.[15-17] SWCNTs grown on $SiO_2$ show almost no detectable PL, which was attributed to the interactions with the polar substrate but also strong nanotube-substrate interactions during growth and mechanical strain.[18] For detailed investigations of individual nanotube spectra, especially at low temperatures, SWCNT are typically embedded in a non-polar polymer matrix.[19-20] However, even in these cases PL lifetime shortening was observed compared to air-suspended SWCNTs[21] and this approach is not applicable for nanotube networks in optoelectronic devices.

In addition to their high sensitivity to the environment, luminescence spectra of SWCNTs feature a series of lower-energy sidebands that appear slightly red-shifted from the main $E_{11}$ exciton peak.[22-24] As shown in **Figure 1c**, these emissive sidebands are typically more prominent in PL spectra of nanotube thin films compared to dispersions. Measurements of isotope-labelled SWCNTs led to the assignment of the $G_1$ band to the bright $E_{11}$ exciton coupled to a G phonon, while the $X_1$ band is linked to the K-momentum dark exciton coupled to a D phonon (see **Figure 1d** for a schematic energy level diagram).[23] The $Y_1$ sideband is associated with defect-induced emission either from an extrinsic state or brightened triplet and has shown large intensity variations between different SWCNTs and even for different positions on the same nanotube.[24] These additional emission features broaden the PL spectrum, which makes clear peak assignments more difficult and is detrimental to applications that require narrow linewidths. Clearly, strategies to reduce or suppress these undesired sidebands will be crucial for applying SWCNT networks in light-emitting devices.

To improve the emission properties of SWCNT thin films and to reduce nanotube-substrate interactions, surface passivation presents a powerful approach. Noé *et al.* demonstrated stable PL emission over time with narrow linewidths for single SWCNTs grown on hexagonal boron nitride (*h*-BN) flakes, while nanotubes on $SiO_2$ exhibited large spectral fluctuations.[25] Likewise, Fang *et al.* observed significantly reduced PL quenching and peak broadening for individual SWCNTs on *h*-BN compared to $SiO_2$.[26] High-quality flakes of *h*-BN have found widespread application as supporting and encapsulating layers for 2D materials and van der Waals heterostructures to reduce their PL linewidth and enhance carrier mobility.[27-30] However, the scalability of this approach is limited, and thus, passivation layers such as polymers that can be rapidly applied over large areas are preferable for the large-scale fabrication of optoelectronic devices. Very recently, the passivation of $SiO_2$ with a cyclic olefin copolymer resulted in high-performance $MoS_2$ field-effect transistors with low interfacial trap densities and high photoresponsivity.[31]

Here, we introduce the polymer BCB (cross-linked B-staged divinyltetramethylsiloxane-bis-benzocyclobutene, see molecular structure in **Figure 1e**) as an ideal passivation layer to enhance the luminescence spectra of SWCNT thin films on commonly used substrates. BCB is a commercially available (as Cyclotene™ resin), low dielectric constant material for microelectronics applications that can be easily and reproducibly processed into thin films with high planarity and optical clarity, low levels of ionic impurities and excellent chemical resistance. The surface of cross-linked BCB has no polar terminal groups or dangling bonds. Its overall structure (aromatic units, mostly C-C and C-H bonds) is very apolar and hydrophobic leading to low water adsorption and absorption. Unlike other polymeric buffer layers, it is insoluble in typical solvents after cross-linking and can withstand photolithography and further processing steps that are required for device fabrication.[32] Due to these excellent properties, BCB has been applied widely as an interlayer dielectric in the industrial fabrication of integrated circuits and multichip modules.[33]

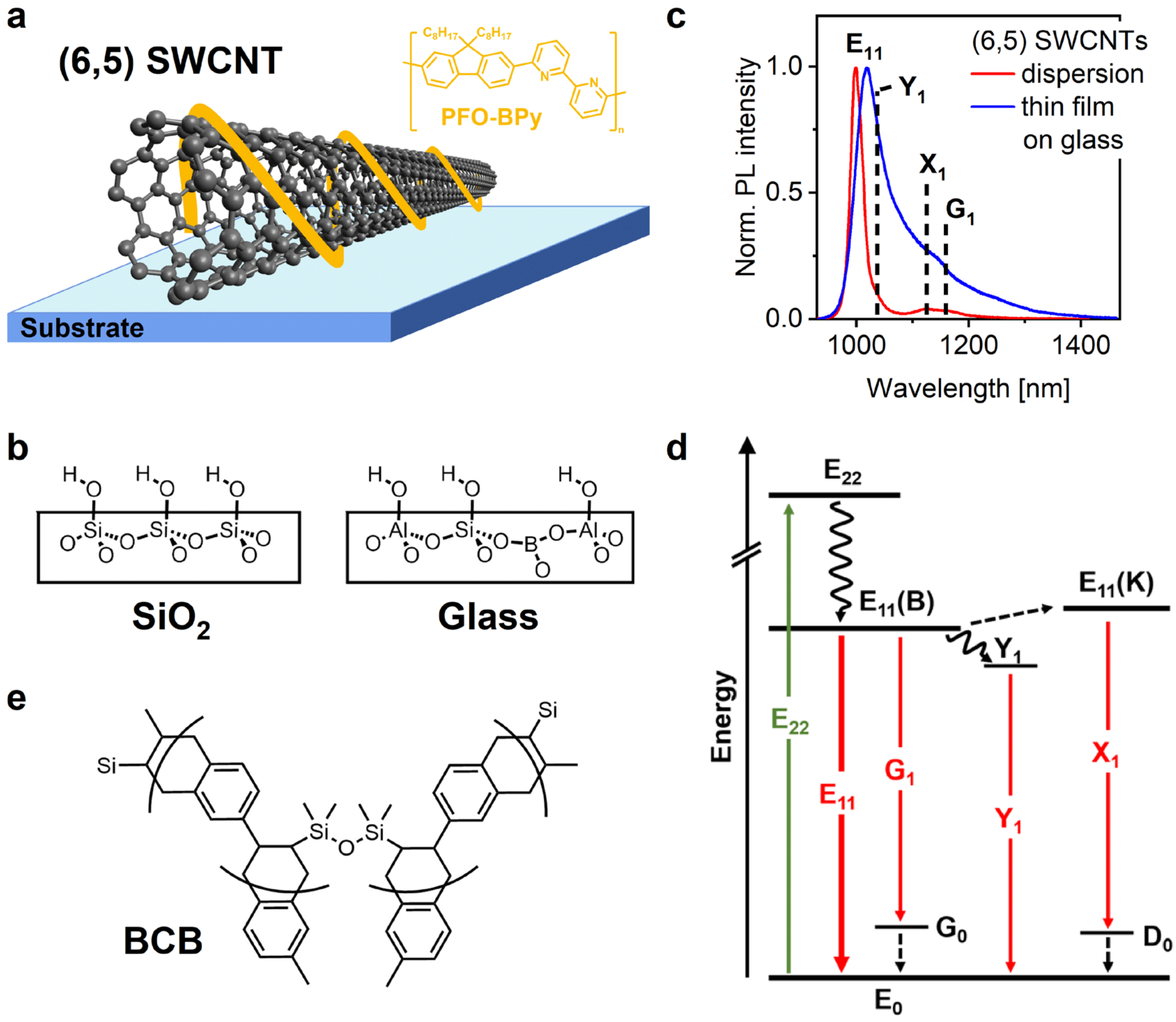

**Figure 1.** **(a)** Schematic illustration of a PFO-BPy-wrapped (6,5) SWCNT on a substrate. The inset shows the molecular structure of the wrapping polymer. **(b)** Schematic drawings of the polar surfaces of thermally-grown $SiO_2$ on silicon and aluminum borosilicate glass as used in this work. **(c)** Normalized PL spectra of (6,5) SWCNTs in a toluene dispersion and a spin-coated thin film on a glass substrate. The $E_{11}$ exciton peak and commonly observed PL sidebands are indicated. **(d)** Energy level diagram (energy axis not to scale) and radiative transitions (red arrows) for semiconducting SWCNTs upon $E_{22}$ excitation (green arrow). Transitions involving phonons are indicated as dashed lines. $E_{11}$(B) is the bright, zero-momentum exciton, $E_{11}$(K) denotes the K-momentum dark exciton. In emission spectra (see Figure 1c), the main $E_{11}$ transition is accompanied by weaker, red-shifted sidebands ($Y_1$, $X_1$, $G_1$). **(e)** Molecular structure of the divinyltetramethylsiloxane-bis-benzocyclobutene based polymer (BCB) after cross-linking.

Notably, the use of BCB as an additional gate dielectric layer prevents electron trapping in back-gated organic field-effect transistors (FETs) much better than different alkyl-terminated self-assembled monolayers, and thus enabled the observation of electron transport in a wide range of organic semiconductors.[34] We find that substrate surface passivation with BCB results in a significant reduction of unwanted PL sideband emission in the luminescence spectra of SWCNT networks, especially of the $Y_1$ sideband, which thus far has been of unknown origin. Our results provide evidence that this $Y_1$ band can be attributed to nanotube lattice defects, that act as shallow exciton traps. They might be introduced through interactions of nanotubes with the substrate surface (*e.g.*, oxygen-containing terminal groups, see **Figure 1b**). While *h*-BN flakes are equally capable of suppressing or preventing the introduction of such defects, we demonstrate the superiority of BCB for homogeneous, large-area surface passivation through fabrication of light-emitting SWCNT network FETs on BCB-treated substrates with significantly narrowed electroluminescence spectra. This passivation concept is also applicable to networks of chemically modified nanotubes with luminescent aryl $sp^3$ defects that were shown to enhance the optical properties of SWCNTs for a variety of applications.[35-37]

## 2. Results and Discussion

To study the impact of nanotube-substrate interactions on the luminescence properties of SWCNT thin films, we used purely semiconducting (6,5) SWCNTs that were obtained *via* exfoliation through shear-force mixing and selective polymer wrapping with the fluorene-bipyridine copolymer PFO-BPy in toluene (for details see Methods; see **Figure 1a** for molecular structure).[5] **Figure S1** (Supporting Information) shows an absorption spectrum of a SWCNT dispersion with a low content of excess, unbound polymer and no other nanotube species. Repeated spin-coating of such dispersions was used to create thin films of SWCNTs on alkali-free aluminum borosilicate glass or Si/$SiO_2$ substrates. PL spectra under resonant excitation at the $E_{22}$ transition show that the characteristic $E_{11}$ exciton emission in thin films (~1018 nm) is red-shifted compared to the corresponding dispersion (~999 nm, see **Figure 1c**). Furthermore, emission from lower-energy PL sidebands ($Y_1$, $X_1$, $G_1$), which is weak in dispersion, becomes rather pronounced in thin films.

The fabrication of optoelectronic devices based on SWCNT networks or thin films typically involves one or several annealing steps, *e.g.*, to remove residual water and oxygen that are detrimental to charge transport and cause undesired doping.[38] Ideally, these procedures should have no adverse impact on the optical properties of the SWCNTs as they were shown to be stable in air up to 400 °C by thermogravimetric analysis.[39] We find however that annealing of spin-coated (6,5) SWCNT films on glass substrates at 150 °C for 30 min already leads to a significant broadening of the $E_{11}$ peak and an increase in sideband emission, as shown in **Figure 2a**. For higher annealing temperatures of 300 °C, emission from the $Y_1$ band becomes significantly stronger, resulting in a shift of the apparent peak maximum to ~1040 nm. Note that the PL spectra in this study typically represent averaged data from more than 100 individual measurements by mapping an area of 50×50 $\mu m^2$. Due to the good homogeneity of films created by spin-coating, the data acquired by this method should be representative for the whole sample. Since thermal annealing was performed in a nitrogen-filled glovebox, we can exclude the effects of oxygen or moisture from ambient air. Instead, we hypothesize that the interaction of SWCNTs with the substrate surface plays a key role for the appearance of these PL sidebands.

To investigate this hypothesis, we passivated glass substrates with a layer of cross-linked BCB polymer (molecular structure shown in **Figure 1e**), which has a low static dielectric constant ε (ε(BCB) = 2.65[33]) and no polar side groups or dangling bonds. A single spin-coating step of the BCB resin followed by brief heating at 290 °C to cross-link the monomers yields a homogeneous film that is essentially insoluble in any common organic solvent with minimal swelling. This is a clear advantage over other polymeric buffer layers such as polystyrene, as it allows for the deposition of SWCNT thin films under identical processing conditions as the untreated substrates. No differences in surface coverage of spin-coated SWCNTs on glass and on BCB were observed, as confirmed by atomic force microscopy (AFM). The PL spectra of (6,5) SWCNT films on glass substrates with a ~75 nm BCB layer show significantly reduced $Y_1$ sideband emission, which only appears as a small shoulder of the $E_{11}$ peak for annealing temperatures as high as 300 °C (**Figure 2a**). We note that even in the PL spectra before annealing, sideband emission and $E_{11}$ peak widths were lower for samples with a BCB layer, as shown in the full data set in **Figure S2** (Supporting Information). Similarly, near-intrinsic emission spectra with less sideband emission was observed for (6,5) SWCNT films on $SiO_2$ substrates passivated with BCB (**Figure 2b** and Supporting Information, **Figure S3**). The $E_{11}$ exciton peak was always consistently blue-shifted by ~5-10 nm for SWCNT films on BCB compared to bare substrates, which is in agreement with the lower dielectric constant of BCB compared to glass and silica (ε(AF32eco glass) = 5.1 and ε($SiO_2$) = 3.9). Importantly, BCB shows no absorption bands in the NIR that could compromise the measured spectra, as shown in **Figure S4** (Supporting Information).

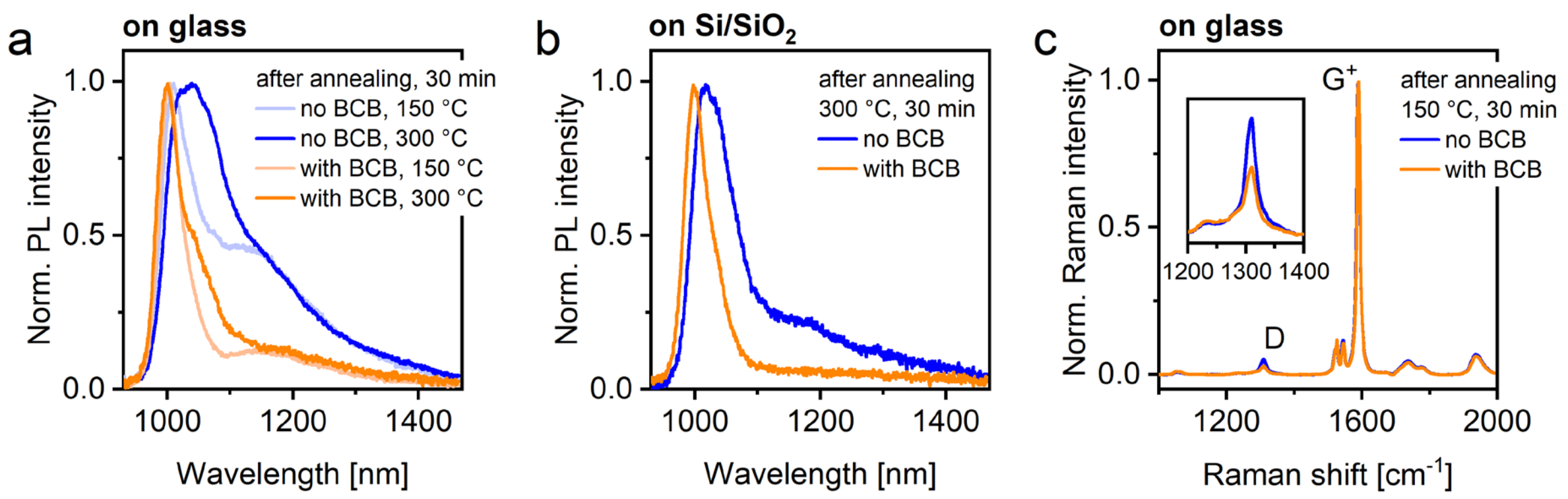

**Figure 2. (a)** Normalized PL spectra of PFO-BPy-sorted (6,5) SWCNT thin films on glass substrates without and with ~75 nm BCB layer after annealing in inert atmosphere at 150 °C and 300 °C for 30 min, respectively. **(b)** Normalized PL spectra of (6,5) SWCNTs on Si/$SiO_2$ substrates without and with ~65 nm BCB layer after annealing in inert atmosphere at 300 °C for 30 min. **(c)** Raman spectra (normalized to the $G^+$ mode) of (6,5) SWCNTs on glass substrates without and with ~75 nm BCB layer after annealing in inert atmosphere at 150 °C for 30 min. The inset shows a zoom-in on the defect-related D mode.

We further tested the influence of the BCB layer thickness as well as the SWCNT network density on the appearance of emissive sidebands in the PL spectra. For BCB layer thicknesses of ~35 nm and ~75 nm, respectively, the resulting spectra were very similar (Supporting Information, **Figure S5**). While the $E_{11}$ peak was only slightly sharper for the thicker BCB layer, both samples showed less sideband emission and $E_{11}$ peak broadening compared to films on non-passivated substrates. A previous study found that BCB layers with a thickness as low as 10 nm still exhibited high dielectric breakdown strengths[32] and thus should also be sufficient to enhance the SWCNT emission spectra. Likewise, the favorable effect of BCB was found irrespective of the density of the SWCNT network (for AFM images and PL spectra, see the Supporting Information, **Figure S6**). Sideband emission was suppressed for dense and sparse nanotube networks on BCB with essentially identical spectra before and after annealing. This observation might be explained with the fast diffusion of excitons along individual nanotubes and within networks toward possible low energy states. Notably, without BCB, the dense films showed more pronounced $X_1$ and $G_1$ sidebands. Since these two sidebands originate from exciton-phonon coupling, this might be due to the closer contact of nanotubes and consequently stronger tube-tube interactions in dense networks.

Deeper insights into the effect of the BCB passivation layer on the (6,5) SWCNT films were obtained from resonant Raman spectroscopy (532 nm excitation wavelength). The Raman spectra in **Figure 2c** show the $G^+$ mode at ~1590 $cm^{-1}$, whose intensity is proportional to the number of $sp^2$-hybridized carbon atoms, as well as the D mode at ~1310 $cm^{-1}$, which is related to the number of defects in the SWCNT lattice.[40] Similar to the PL data, the Raman spectra were averaged over more than 2500 individual measurements from a 100×100 $\mu m^2$ map to rule out any differences due to local inhomogeneities. Clearly, the intensity of the D mode relative to the $G^+$ mode after annealing at 150 °C is higher for the sample without BCB compared to the BCB sample. Importantly, for the as-prepared films before annealing, the $D/G^+$ peak area ratio was almost identical (Supporting Information, **Figure S7** and **Table S1**). Subsequent annealing led to an increase of the $D/G^+$ ratio for the bare glass substrate while it remained essentially the same for substrates with BCB passivation. Measurements on SWCNT films on $Si/SiO_2$ without and with BCB gave the same results (Supporting Information, **Figure S8** and **Table S2**). The observation of a higher defect-related D mode indicates that the increase in emission from lower-energy sidebands is correlated with a larger number of lattice defects in the SWCNTs. Very recently, Larson *et al.* investigated oxygen-mediated photodegradation in PFO-BPy-wrapped (6,5) SWCNTs and found a substantial increase in the Raman D mode as well as a drop in PL intensity and $E_{11}$ peak broadening.[41] A red-shift and broadening of the $E_{11}$ emission was also observed for guanine defects that are photochemically introduced in DNA-wrapped SWCNTs.[42-43] Hence, we infer that the $Y_1$ band, whose origin has remained elusive thus far, might be associated with structural defects in the SWCNT sidewalls. According to a recently developed method to quantify the number of $sp^3$ lattice defects in small-diameter SWCNTs *via* Raman spectroscopy,[44] the observed increase in the $D/G^+$ ratio (from 0.08 to 0.11) for the sample without BCB upon annealing would correspond to ~12 introduced defects per µm of SWCNT.

Notably, the formation of $sp^3$-hybridized defects at oxide interfaces was previously observed for graphene on which aluminum oxide was deposited by electron beam evaporation.[45] Similarly, electron beam deposition of aluminum oxide and $SiO_2$ on individual SWCNTs gave rise to exciton-localizing, luminescent oxygen defects with red-shifted emission wavelengths[46-47] that even behaved as single-photon emitters.[48] We thus hypothesize that the controlled introduction of red-shifted, light-emitting oxygen defects might also be achieved using reactive oxidic surfaces. Previous reports of oxygen functionalization of SWCNTs in dispersion mainly involved light irradiation in the presence of ozone or hypochlorite.[49-51] The resulting variety of red-shifted emission features was attributed to different configurations of the oxygen adducts, *i.e.*, ether and epoxide defects.[50] Importantly, low-temperature PL spectra showed that oxygen functionalization also leads to additional peaks appearing in the region of $E_{11}$ emission, most notably a transition (termed $E_{11}^{-}$) that was red-shifted by ~26 meV and behaved similar to the deeper trap states ($E_{11}$*).[50] These results suggest that the $Y_1$ sideband might be of similar origin. Here, we must emphasize that the inferred role of defects in the emergence of the $Y_1$ band is different from the recently proposed population mechanism of the K-momentum dark excitonic state (which gives rise to the $X_1$ sideband emission feature, see **Figure 1d**) through exciton scattering at lattice defects.[52-53]

To further corroborate that the $Y_1$ band is associated with lattice defects, we performed low-temperature PL spectroscopy to determine the temperature dependence of sideband emission intensities. **Figure 3a** and **3b** show the PL spectra of (6,5) SWCNT networks measured on a bare glass substrate and on a substrate passivated with a BCB layer between 100 – 300 K. In agreement with our previous observations, the spectrum for the sample on glass at 300 K (orange curve in **Figure 3a**) shows a broad tail of the $E_{11}$ emission peak extending beyond 1300 nm, which is much less pronounced for the film on BCB. Decreasing the temperature resulted in an increase in sideband emission at ~1070 nm ($Y_1$ band) that eventually becomes dominant over the $E_{11}$ emission feature. An increase of the $Y_1$ band was also found for the sample with BCB, although it was much less significant. Notably, below 100 K, no further change in the spectral shape was observed for either sample (Supporting Information, **Figure S9**). Consistent with the data obtained by Raman spectroscopy, the rise of the $Y_1$ sideband emission relative to the $E_{11}$ at lower temperatures indicates an energetically lower-lying defect state as the origin. To quantify the energetic difference, PL spectra were fitted with three Gaussian peaks ($E_{11}$, $Y_1$, and a third contribution to account for the tail at longer wavelengths). From linear fits of the logarithmic $Y_1/E_{11}$ peak area ratio *versus* inverse temperature (**Figure 3c**), we obtain very similar thermal trap depths of ~40 meV and ~32 meV for the data set without BCB and with BCB, respectively. Such shallow energetic traps would allow for fast de-trapping of excitons at room temperature (thermal energy $k_B T \approx 25$ meV) that is slowed down at lower temperatures. These results are consistent with those previously reported for cryogenic PL spectra of individual, untreated (6,5) SWCNTs, which revealed the spontaneous localization of excitons in shallow trap states that were attributed to covalent sidewall defects.[54] The trap depth values are also similar to those found for intentionally introduced, luminescent alkyl or aryl $sp^3$ defects (thermal trap depths of ~75-150 meV for $E_{11}$* defects and ~25 meV for $E_{11}^{*-}$ defects) or oxygen ether defects (~24 meV).[55-56] Note that the thermal trap depths of luminescent defects are different from the optical trap depths, *i.e.*, their spectral red-shift.[55] Moreover, power-dependent PL spectra under pulsed $E_{22}$ excitation and non-resonant continuous-wave excitation showed a relative decrease in sideband to $E_{11}$ emission intensities with increasing excitation power (Supporting Information, **Figure S10**) as reported previously for oxygen and $sp^3$ aryl defects (state-filling effect).[57] These findings further corroborate that shallow excitonic trap states are the most likely origin of the $Y_1$ sideband.

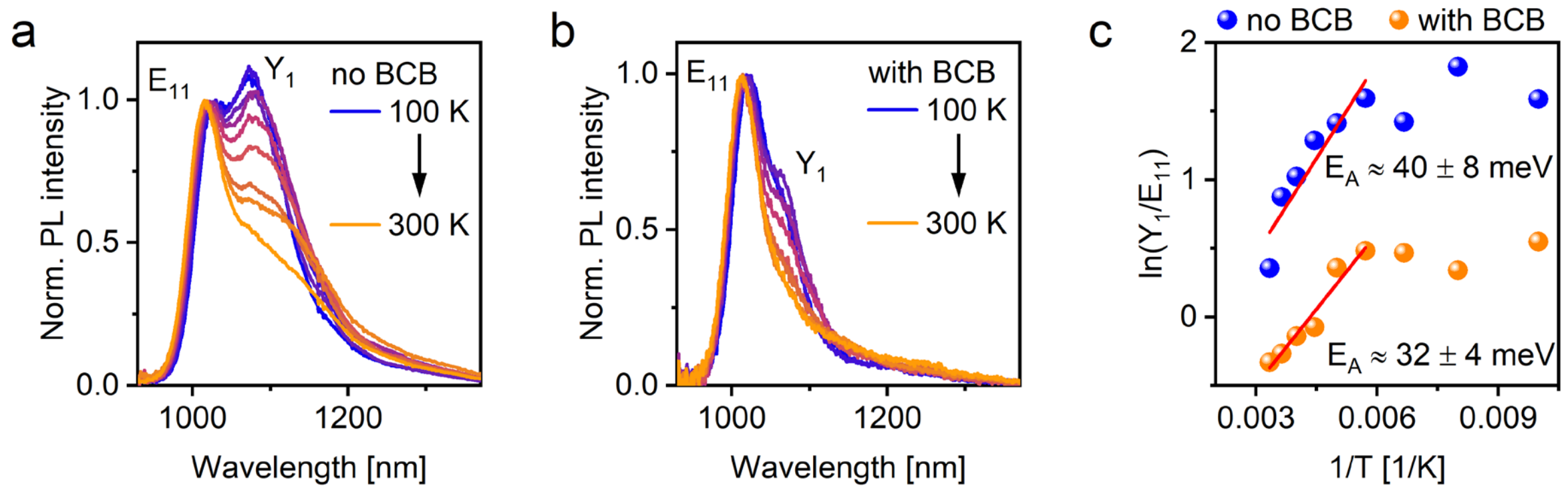

**Figure 3.** Normalized temperature-dependent PL spectra of (6,5) SWCNT thin films on glass substrates **(a)** without and **(b)** with ~80 nm BCB layer between 100 K and 300 K in steps of 25 K. Samples were annealed in inert atmosphere at 150 °C for 45 min. **(c)** Logarithmic peak area ratios of $Y_1$ and $E_{11}$ peaks obtained from fits of temperature-dependent PL spectra without and with BCB *versus* inverse temperature. Red lines are linear fits to the data between 175 K and 300 K.

Due to their well-established sorting process, (6,5) SWCNTs are commonly used and emission from PL sidebands in their spectra have been frequently observed.[22-23] Only few studies however utilized thin films composed of different monochiral SWCNTs for spectroscopic studies. Jakubka *et al.* reported photo- and electroluminescence from electrolyte-gated transistors with (6,5), (7,5) and (10,5) SWCNT networks.[11] While the PL spectrum of the small-diameter (6,5) SWCNTs showed pronounced sideband emission, it was virtually absent in thin films of larger-diameter (7,5) and (10,5) SWCNTs. This observation raises the question, which parameters are necessary for the occurrence of PL sidebands and how generalizable the surface passivation with BCB is. To this end, we prepared dispersions of (7,5) SWCNTs wrapped with the polyfluorene derivative PFO (for details see Methods). Indeed, the spectra of (7,5) SWCNT networks on glass substrates without and with BCB layer showed very little emission from PL sidebands even after annealing at 300 °C (Supporting Information, **Figure S11**). The $E_{11}$ peak was slightly narrower and blue-shifted by ~4 nm for thin films on BCB. Prolonged annealing of (7,5) SWCNT networks on $SiO_2$ substrates at 300 °C eventually resulted in the appearance of the $Y_1$ band but could once more be circumvented by applying a BCB layer, as shown in **Figure 4a** (full data set shown in the Supporting Information, **Figure S12**).

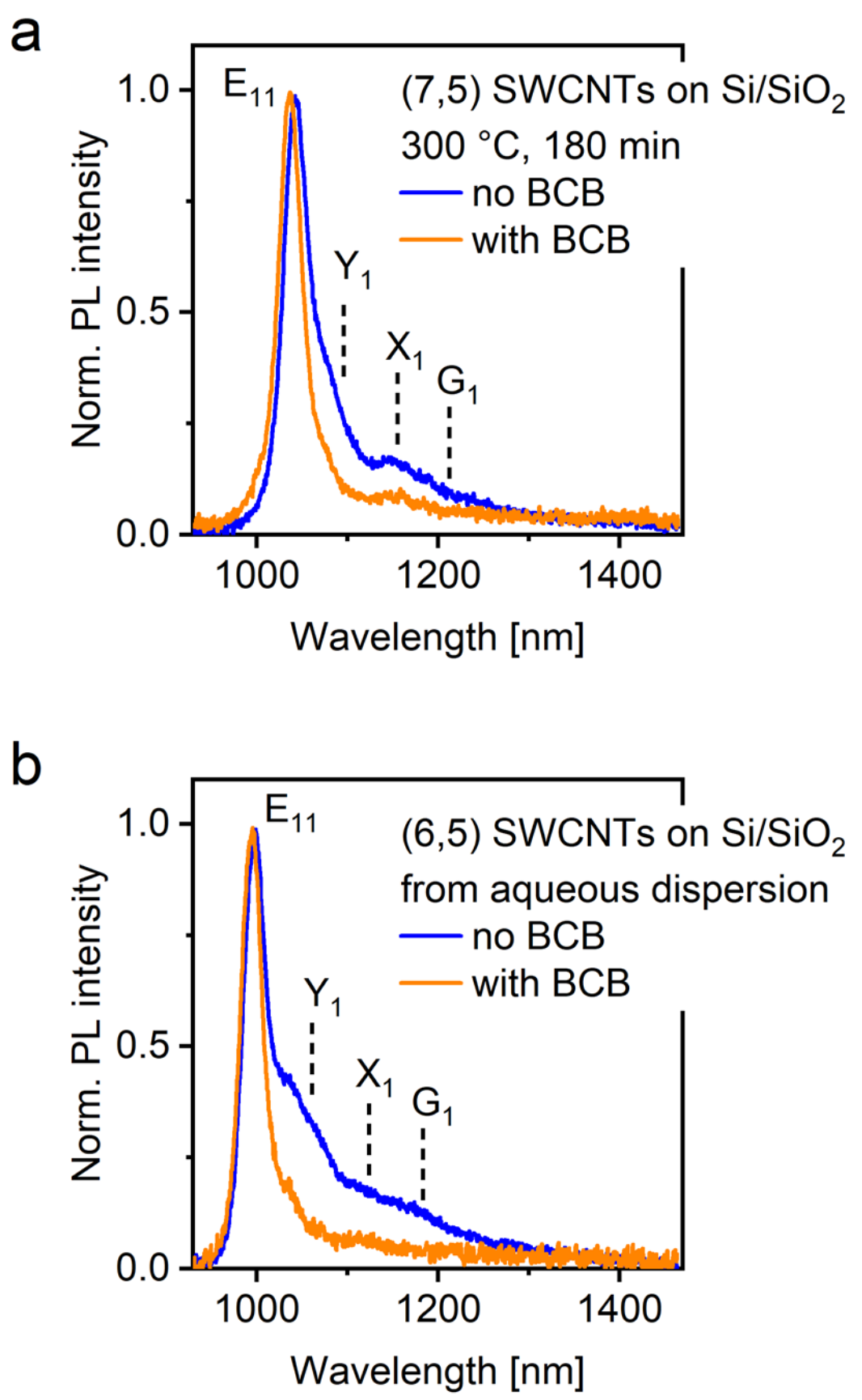

**Figure 4. (a)** Normalized PL spectra of (7,5) SWCNT thin films on $Si/SiO_2$ substrates without and with ~75 nm BCB layer after annealing in inert atmosphere at 300 °C for 180 min. **(b)** Normalized PL spectra of (6,5) SWCNT thin films deposited from aqueous dispersion with sodium deoxycholate on $Si/SiO_2$ substrates without and with ~75 nm BCB layer.

An explanation for the discrepancy between (6,5) and (7,5) SWCNTs could involve their different geometry. Due to their larger diameter, (7,5) SWCNTs are less strained and therefore less reactive, which has been observed previously for the chemical introduction of covalently bound $sp^3$ defects.[44,56] However, to rule out that the observed effects are related to the wrapping polymer (with or without bipyridine repeat units), we also used dispersions of (6,5) SWCNTs stabilized with sodium deoxycholate in water that were sorted by the aqueous two-phase extraction method.[6] The PL spectra of spin-coated films of these (6,5) SWCNTs on Si/$SiO_2$ in **Figure 4b** unambiguously show sideband emission compared to the narrow, near-intrinsic $E_{11}$ emission for equivalent networks on a BCB layer, thus confirming that the observed sidebands indeed originate from nanotube-surface interactions and not from polymer-specific interactions.

A prominent material for surface passivation and encapsulation of low-dimensional materials is hexagonal boron nitride (*h*-BN). It has been widely used for 2D materials and recently also for individual SWCNTs.[25-28] To elucidate the effect of *h*-BN on the spectra of SWCNT networks and to compare it to the effects of BCB, we transferred multiple *h*-BN flakes of several tens of micrometers in lateral size and several tens of nanometers in thickness (static dielectric constants of *h*-BN: ε(in-plane, bulk) = 4.98, ε(out-of-plane, bulk) = 6.93)[58] onto Si/$SiO_2$ substrates by mechanical exfoliation. An optical microscopy image of a representative *h*-BN flake is shown in **Figure 5a**. Thin films of (6,5) SWCNTs were prepared by spin-coating to yield nanotube networks with similar densities on the *h*-BN flakes and on the bare substrate as confirmed by AFM. Hyperspectral PL imaging revealed a significantly higher integrated $E_{11}$-to-sideband PL ratio (*i.e.*, less sideband emission relative to the main $E_{11}$ excitonic emission peak) on the *h*-BN compared to the $SiO_2$ surface, as shown in **Figure 5b**. The outline of the *h*-BN flake is clearly visible when comparing the PL map to the optical micrograph, which confirms that *h*-BN just as BCB prevents nanotube-substrate interactions that would lead to the appearance of PL sidebands. PL spectra from two representative spots (positions indicated by circles in **Figure 5b**) also show a significantly sharper $E_{11}$ peak and significantly reduced sideband emission on the *h*-BN flake in comparison to measurements on the bare substrate (**Figure 5c**). Overall, the PL spectra of (6,5) SWCNTs on *h*-BN are very similar to those on BCB. Additional PL spectra on Si/$SiO_2$ substrates and on several different *h*-BN flakes together with optical micrographs are presented in **Figure S13** (Supporting Information). Raman mapping of SWCNTs on *h*-BN flakes also showed a substantially lower D/$G^+$ ratio on the flake compared to the bare substrate before and after annealing (Supporting Information, **Figure S14**), further corroborating a lower number of surface-induced defects.

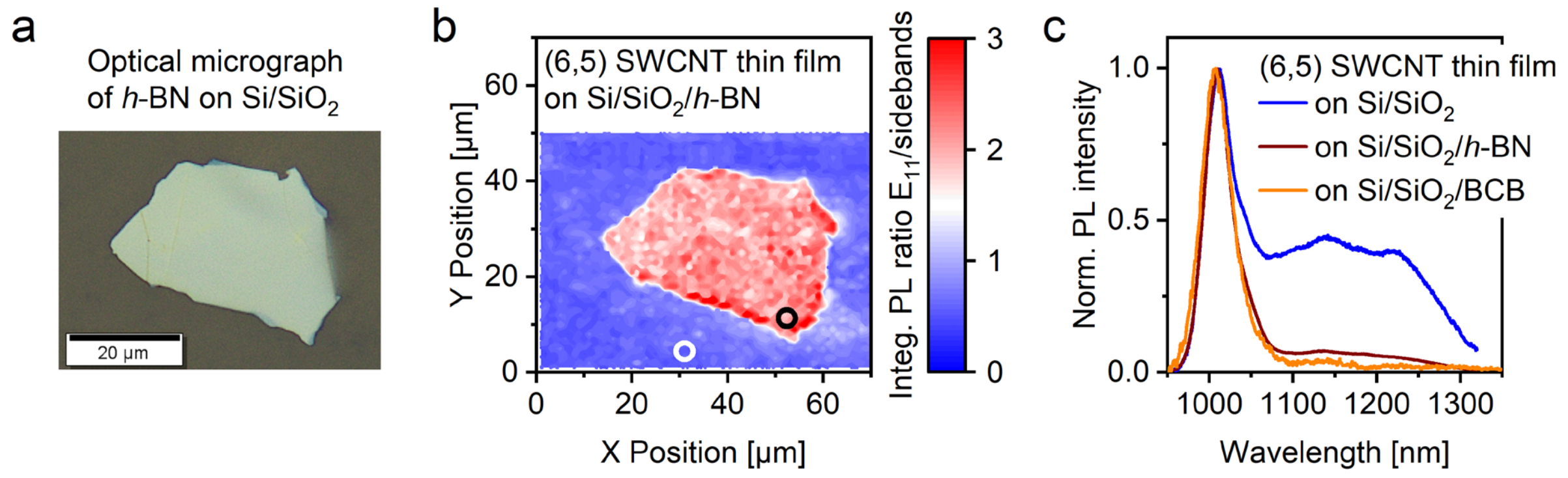

**Figure 5. (a)** Optical micrograph of a *h*-BN flake on $Si/SiO_2$. **(b)** PL map showing the ratio of the integrated $E_{11}$-to-sideband PL intensity (integration from 950-1050 nm and 1100-1300 nm, respectively) of a (6,5) SWCNT thin film on a *h*-BN flake on $Si/SiO_2$. **(c)** Normalized PL spectra on the substrate (blue line) and on the *h*-BN flake (brown line) recorded from the spots indicated by circles in (b). The orange line shows a normalized PL spectrum on $Si/SiO_2$/BCB for comparison.

While BCB and *h*-BN are equally effective to prevent the formation of defects and sideband emission in PL spectra of (6,5) SWCNT thin films, the advantages of the BCB passivation layer include its facile application as a homogeneous film over large areas, its compatibility with typical processing steps used in the fabrication of optoelectronic devices,[59] and its commercial availability. To highlight these benefits, we fabricated (6,5) SWCNT network top-gate field-effect transistors (FETs) on glass substrates with ~80 nm BCB. **Figure 6a** shows a schematic cross-section of a FET with interdigitated gold source/drain electrodes (channel length $L$ = 20 µm, channel width $W$ = 10 mm), dense (6,5) SWCNT networks, a double-layer PMMA/hafnium oxide ($HfO_x$) dielectric, and a silver top gate electrode. AFM images of SWCNT networks are shown in **Figure S15** (Supporting Information). Neither the photolithography steps nor electron beam evaporation to deposit the metal contacts, lift-off in *N*-methyl pyrrolidone, nor the short oxygen plasma treatment to pattern the SWCNT networks showed any detrimental effects on the BCB layer.

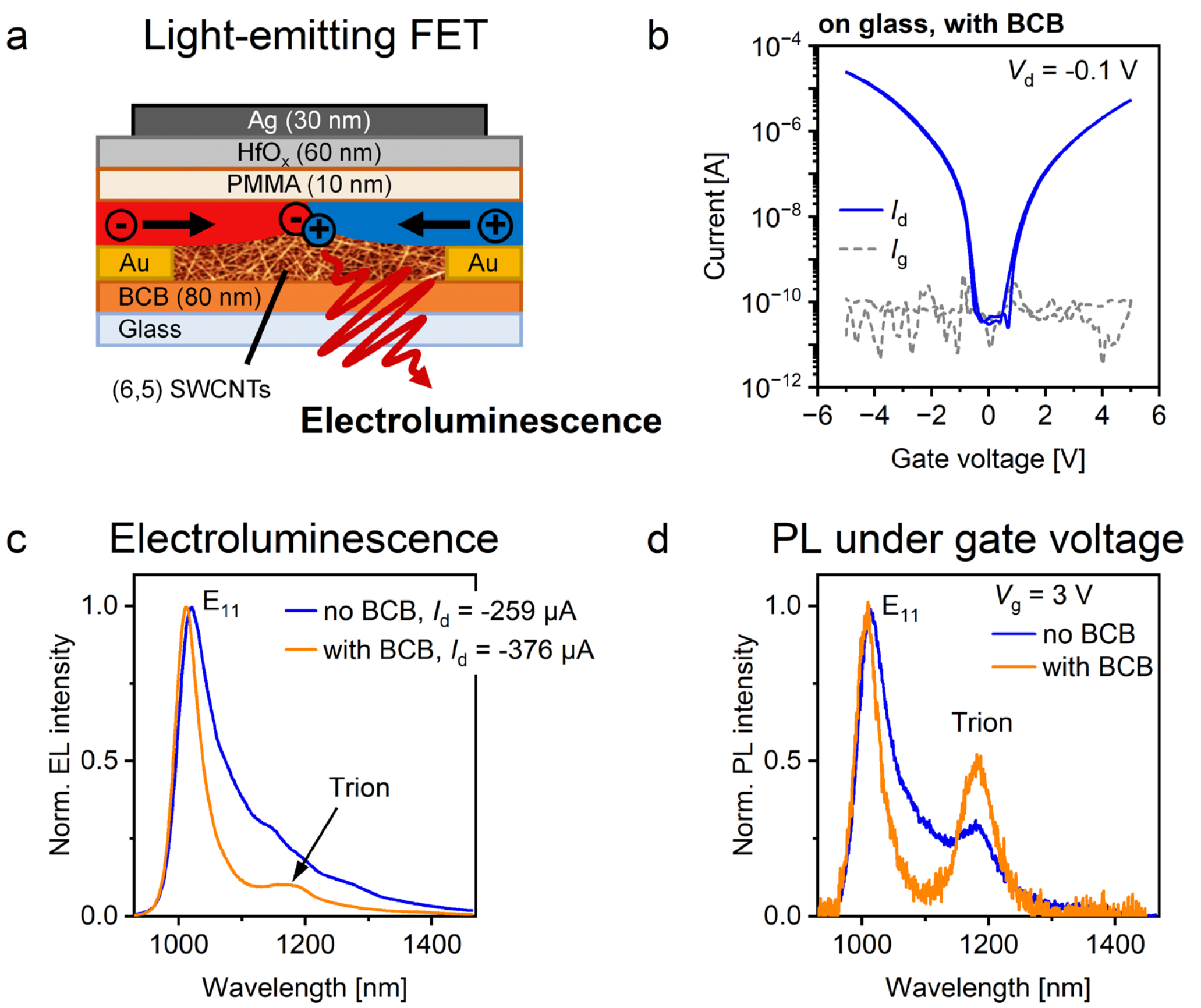

**Figure 6. (a)** Schematic cross-section of a light-emitting (6,5) SWCNT network FET on a glass substrate with ~80 nm BCB layer (layer thicknesses not to scale). Upon accumulation of electrons (red) and holes (blue) in the ambipolar regime, carrier recombination gives rise to electroluminescence from a narrow emission zone in the FET channel. **(b)** Transfer characteristics of a (6,5) SWCNT network FET on glass with BCB in the linear regime ($V_d$ = -0.1 V). The blue solid line is the drain current $I_d$, the gray dashed line is the gate leakage current $I_g$. **(c)** EL spectra from SWCNT FETs on glass without and with BCB layer at similar drain currents $I_d$ normalized to the $E_{11}$ exciton emission. **(d)** PL spectra measured under applied gate voltage $V_g$ at similar intensities of trion emission (~1180 nm) normalized to the $E_{11}$ exciton emission. Samples in **(b-d)** were annealed in inert atmosphere at 150 °C for 45 min.

Irrespective of the presence or absence of the passivation layer, all FETs showed balanced ambipolar charge transport (*i.e.*, holes and electrons) as evident from the linear transfer characteristics (**Figure 6b** and Supporting Information, **Figure S16**). Output curves and transconductance data are shown in **Figure S17** (Supporting Information). Averaged charge carrier mobilities for holes and electrons remained essentially the same for samples with BCB or without (Supporting Information, **Table S3**). In a previous study, BCB had enabled the observation of electron transport in semiconducting polymers that was otherwise prevented by electron trapping at the hydrophilic $SiO_2$ dielectric interface.[34] Another recent report demonstrated significantly improved electrical characteristics of $MoS_2$ FETs passivated with a cyclic olefin copolymer.[31] However, these studies employed a bottom-gate FET architecture on $Si/SiO_2$ substrates, in which charge transport occurs at the interface between semiconductor and the $SiO_2$ dielectric. In the bottom-contact/top-gate device structure used here, the electrostatic field confines charge accumulation and transport to the interface between SWCNT network and the dielectric layer (here PMMA). Consequently, charge carriers do not directly interact with the substrate surface and hence, no large impact of surface passivation on the carrier mobilities is expected.

When operated in the ambipolar regime, the simultaneous accumulation of electrons and holes in the FET channel gives rise to a narrow recombination and emission zone[10] and consequently electroluminescence (EL) as schematically shown in **Figure 6a**. The emission intensity is determined by the current flow between source and drain electrodes (drain current $I_d$). EL spectra of SWCNT FETs without and with BCB acquired at similar drain currents are presented in **Figure 6c**. Clearly, the BCB layer leads to near-intrinsic luminescence with lower sideband emission and a sharper, blue-shifted $E_{11}$ peak not only for PL but also in EL spectra. It also allows us to identify the lower-energy trion (charged exciton) emission peak at ~1165 nm[11] that emerges for high drain currents. Even though the impact of substrate passivation on charge transport in these devices is not significant, the improvement of the EL spectra is striking. This can again be attributed to the high mobility of excitons in individual nanotubes and in dense networks,[60-62] as they are not confined to the semiconductor/dielectric interface by the electric field as opposed to the charge carriers. As such, excitons are much more sensitive to even few defect sites within the SWCNT network and the nanotube-substrate interface.

A complete data set of EL spectra for different drain currents is shown in **Figure S18** (Supporting Information). For light-emitting FETs with conventional dielectrics (*i.e.*, not electrolyte-gated), trion emission is typically low in EL spectra but becomes more pronounced in PL experiments conducted under static doping by applying a fixed gate voltage.[63] As shown in **Figure 6d**, the use of a passivating BCB layer gives rise to a clean spectrum with two distinct peaks corresponding to exciton and trion emission, respectively. Hence, this approach enables an unambiguous and detailed analysis of the trion peak, which was previously complicated by $E_{11}$ sidebands.[64] For gate voltage-dependent PL spectra in hole and electron accumulation regimes for samples without and with BCB, see the Supporting Information, **Figure S19** and **S20**.

Aside from trions, luminescent *sp*$^3$ defects are another source of red-shifted emission in SWCNTs and have recently attracted great interest due to their tunable emission wavelengths and ability to act as room-temperature single-photon emitters.[35-37] To demonstrate the applicability of BCB passivation to nanotube films with *sp*$^3$ defects, we covalently functionalized (6,5) SWCNTs with diazonium salts to introduce $E_{11}$* defects (emission wavelength ~1180 nm) and with aniline derivatives to yield $E_{11}$*$^-$ defects (emission wavelength ~1250 nm) as shown previously.[56,65] For details regarding the functionalization procedures see the Supporting Information. Again, PL spectra of spin-coated SWCNT films on glass showed less unwanted sideband emission for BCB-passivated substrates, but sharper $E_{11}$ and *sp*$^3$ defect emission peaks (Supporting Information, **Figure S21** and **S22**). We further conducted time-resolved PL measurements at the wavelengths of defect emission to obtain insights into their PL dynamics in thin films on BCB-passivated and non-passivated glass substrates. In SWCNT dispersions, defect-localized excitons exhibit significantly longer PL lifetimes compared to mobile $E_{11}$ excitons as the latter are prone to diffusive quenching.[65] Note that the lifetime of the mobile $E_{11}$ excitons is only few ps and therefore instrument-limited in our setup. When processed into thin films, (6,5) SWCNTs with *sp*$^3$ defects also show significant lifetime shortening (Supporting Information, **Table S4** and **S5**). However, for the PL decays measured on the BCB-treated substrates, we find substantially longer defect state lifetimes indicating reduced non-radiative exciton decay probably due to the lower dielectric constant of the environment.[21] This effect opens up new avenues for the application of *sp*$^3$-functionalized SWCNTs as tunable, solid-state narrowband NIR emitters.[63] Overall, BCB emerges as an ideal substrate passivation layer for light-emitting SWCNT thin films and devices.

## 3. Conclusion

In summary, we have investigated the effect of surface passivation with the commercially available, low dielectric constant, cross-linked polymer BCB on the spectroscopic properties of semiconducting SWCNT thin films. We find that PL spectra of SWCNT networks on BCB-treated substrates (such as glass or Si/$SiO_2$) show significantly reduced emission from lower-energy sidebands, especially the so-called $Y_1$ band, which often leads to detrimental spectral broadening. Raman and temperature-dependent PL spectra indicate that the observed $Y_1$ sideband emission is related to structural defects in the SWCNT lattice that are likely introduced by interaction of the nanotube lattice with oxygen-containing terminal groups on glass or $SiO_2$ surfaces. The surface reactivity of common substrates should not be ignored and adds another important parameter beyond their dielectric constant. While flakes of *h*-BN provide a similar level of passivation and improved PL spectra, the easy and reproducible fabrication of large-area, homogeneous and highly planar BCB films by a simple spin-on and annealing procedure is highly advantageous. Combined with its ability to withstand most solvents and chemicals, BCB is ideally suited as a high-quality surface passivation layer. The realization of ambipolar, light-emitting SWCNT network field-effect transistors on BCB-coated glass substrates further demonstrates its excellent compatibility with typical device fabrication and processing steps. In summary, we propose BCB as a standard surface coating layer for optoelectronic devices with, and spectroscopic investigations of, SWCNTs as well as low-dimensional emitters such as monolayers of transition metal dichalcogenides and other layered semiconductors.

## 4. Experimental Section

*Selective Dispersion of (6,5) and (7,5) SWCNTs*: Toluene-based dispersions of (6,5) and (7,5) SWCNTs were prepared from CoMoCAT raw material (Sigma Aldrich, batch MKCJ7287, 0.4 g $L^{-1}$) *via* selective polymer-wrapping and shear force mixing (Silverson L2/Air mixer, 10,230 rpm, 72 h, 20 °C) as previously described in detail.[5] (6,5) SWCNTs were dispersed with poly[(9,9-dioctylfluorenyl-2,7-diyl)-*alt*-(6,6'-(2,2'-bipyridine))] (PFO-BPy, American Dye Source, $M_W$ = 40 kg $mol^{-1}$, 0.5 g $L^{-1}$), whereas (7,5) SWCNTs were dispersed with poly(9,9-dioctylfluorene) (PFO, American Dye Source, $M_W$ > 20 kg $mol^{-1}$, 0.9 g $L^{-1}$). After exfoliation, the resulting dispersions were centrifuged twice at 60,000*g* for 45 min (Beckman Coulter Avanti J26XP centrifuge), followed by filtration of the supernatant through a polytetrafluoroethylene (PTFE) syringe filter (Whatman, pore size 5 µm) to remove residual aggregates and impurities. SWCNTs were collected by vacuum filtration through a PTFE membrane (Merck Millipore JVWP, pore size 0.1 µm) and the filter cakes were thoroughly washed with toluene to reduce the content of excess polymer. The final SWCNT dispersions were obtained by bath sonication of the filter cakes for 30 min in fresh toluene directly prior to use for thin film deposition or $sp^3$ functionalization (for experimental details, see the Supporting Information). Aqueous dispersions of (6,5) SWCNTs were prepared from CoMoCAT raw material (CHASM SG65i-L58) by aqueous two-phase extraction (ATPE) as described previously.[6,66] The two-phase system was composed of dextran ($M_W$ = 70 kDa, TCI) and poly(ethylene glycol) (PEG, $M_W$ = 6 kDa, Alfa Aesar), and SWCNTs were separated according to a diameter sorting protocol based on sodium deoxycholate (DOC, BioXtra) and sodium dodecyl sulfate (SDS, Sigma Aldrich). The separation of metallic and semiconducting SWCNT species was achieved with sodium hypochlorite (NaClO, Sigma Aldrich) as an oxidant. The sorted (6,5) SWCNTs were concentrated using a pressurized ultrafiltration stirred cell (Millipore) with a 300 kDa $M_W$ cutoff membrane and adjusted to 2% (w/v) DOC for further processing.

*Preparation of BCB Layers, h-BN Flakes, and SWCNT Thin Films*: Alkali-free aluminum borosilicate glass substrates (Schott AF32eco, 300 µm thickness) and silicon wafers (500 µm thickness with 300 nm thermally grown $SiO_2$, Siegert Wafer GmbH) were cleaned by ultrasonication in acetone and 2-propanol for 10 min each and transferred into a nitrogen-filled glovebox. The divinyltetramethylsiloxane-bis-benzocyclobutene (BCB) resin precursor (Cyclotene 3022-35, Lot: D191L7D007-070) was diluted with appropriate amounts of mesitylene to give different final layer thicknesses. The diluted solution was spin-coated (3 seconds at 500 rpm, followed by 60 seconds at 8000 rpm) onto glass or silicon substrates and the BCB was cross-linked at 290 °C (2 min). Under these conditions, the degree of cross-linking should be >95%.[32] Flakes of high-quality *h*-BN with different size and thickness were obtained by mechanical exfoliation from bulk crystals grown under high pressure[67] using adhesive tape. The tape was brought into contact with a silicon substrate (525 μm thickness, coated with 80 nm thermally grown $SiO_2$), heated to 110 °C for 3 minutes, and pulled off after cooling to room temperature, leaving the flakes on the substrate. SWCNT networks were deposited on the substrates from toluene-based dispersions under ambient conditions *via* repeated spin-coating (2000 rpm, 30 s, 3-5 times) with short intermediate annealing steps (100 °C, 2 min). To remove excess polymer, substrates were rinsed with THF and 2-propanol. If appropriate, subsequent annealing of SWCNT thin films at different temperatures was carried out in a nitrogen-filled glovebox.

*Fabrication of (6,5) SWCNT Network FETs*: Interdigitated bottom-contact electrodes ($L$ = 20 μm, $W$ = 10 mm) were patterned on glass substrates (Schott AF32eco, 300 μm thickness) with and without BCB layer by photolithography (LOR5B/S1813 resist, microresist technology; Suess MicroTec MA6 mask aligner), followed by electron beam evaporation of chromium (3 nm) and gold (30 nm). Lift-off was performed in *N*-methyl pyrrolidone. Dense networks of (6,5) SWCNTs were deposited from concentrated dispersions (optical density of 8 $cm^{-1}$ at the $E_{11}$ absorption peak) *via* repeated spin-coating (2000 rpm, 30 s) with annealing steps (100 °C, 2 min) in between. Substrates were rinsed with THF and 2-propanol to remove excess polymer before the SWCNT networks were patterned by photolithography and oxygen plasma treatment (Nordson MARCH AP-600/30, 100 W, 2 min) to remove all nanotubes outside the channel area. After annealing in inert atmosphere (150 °C, 45 min), a double-layer dielectric was deposited. First, ~10 nm of poly(methyl methacrylate) (PMMA, Polymer Source, $M_W$ = 315 kg $mol^{-1}$, syndiotactic) were spin-coated (4000 rpm, 60 s) from *n*-butyl acetate (6 mg $mL^{-1}$). Subsequently, atomic layer deposition (Ultratech Savannah S100) was used to deposit ~60 nm of hafnium oxide at 100 °C using a tetrakis(dimethylamino)hafnium precursor (Strem Chemicals) and water as oxidizing agent. Thermal evaporation of 30 nm silver top-gate electrodes through a shadow mask completed the devices.

*Characterization*: Film thicknesses were determined with a Bruker DektakXT Stylus profilometer. Atomic force microscopy (AFM) images of the surface topography under ambient conditions were recorded with a Bruker Dimension Icon in ScanAsyst mode. Baseline-corrected absorption spectra were measured with a Varian Cary 6000i UV-vis-NIR spectrometer. Raman spectra and maps (532 nm excitation, grating 2400 lines $mm^{-1}$) were acquired with a Renishaw inVia confocal Raman microscope in backscattering configuration equipped with a 50× long working distance objective (N.A. 0.5, Olympus). Spectra were averaged from >2500 individual measurements over an area of 100×100 $\mu m^2$. High-resolution Raman maps of the samples were acquired in Streamline mode. Peaks were fitted separately using a combination of a Lorentzian and a Gaussian line shape in the Wire 3.4 software. For the acquisition of photoluminescence (PL) spectra, samples were excited with the output of a picosecond-pulsed supercontinuum laser (NKT Photonics SuperK Extreme) which was wavelength-filtered at 575 nm for (6,5) SWCNTs and 655 nm for (7,5) SWCNTs, respectively. The laser was focused onto the samples and the emission was collected with a NIR-optimized 50× objective (N.A. 0.65, Olympus). Emission spectra were recorded with a Princeton Instruments Acton SpectraPro SP2358 spectrograph (grating 150 lines $mm^{-1}$) equipped with a liquid nitrogen cooled InGaAs line camera (Princeton Instruments OMA V:1024−1.7 LN). Typically, >100 spectra were recorded by mapping an area of 50×50 $\mu m^2$ with a XYZ piezo stage (Nano-LPS200, Mad City Labs) and the signal was averaged. Temperature-dependent PL spectra of (6,5) SWCNT films and room-temperature spectra of (6,5) SWCNTs on *h*-BN flakes were measured in high vacuum (< $10^{-5}$ mbar) under continuous wave laser excitation at 532 nm (OBIS, Coherent) using a closed-cycle liquid helium cooled cryostat (Montana Cryostation s50). The excitation beam was focused and emitted photons were collected with a NIR-optimized 50× objective (N.A. 0.42, Mitutoyo) that was mounted outside the cryostat. PL emission was spectrally resolved with an IsoPlane SCT-320 spectrograph (Princeton Instruments, grating 150 lines $mm^{-1}$) and recorded with a thermoelectrically cooled 2D InGaAs array (NIRvana 640ST). Hyperspectral PL imaging of SWCNTs on *h*-BN flakes was performed in a home-built scanning confocal microscope. The sample was mounted on piezo-stepping units (ANPxyz101, attocube systems) for positioning with respect to the confocal spot of an apochromatic objective (LT-APO/IR/0.81, attocube systems). A Ti:sapphire laser (Mira900, Coherent) in continuous wave mode was used for excitation around 750 nm with a power of 2 µW. The PL was spectrally dispersed by a monochromator (Princeton Instruments Acton SP-2500) and detected with a liquid nitrogen-cooled InGaAs array (Princeton Instruments OMA V:1024−1.7 LN).

**Supporting Information**

Supporting Information is available from the Wiley Online Library or from the author.

**Acknowledgements**

This project has received funding from the European Research Council (ERC) under the European Union's Horizon 2020 research and innovation programme (Grant agreement no. 817494 "TRIFECTs"). The authors are very grateful to T. Taniguchi and K. Watanabe (NIMS, Japan) for providing *h*-BN crystals as a reference material. A. H. acknowledges funding by the European Research Council (ERC) under the European Union's Horizon 2020 research and innovation programme under the Grant agreement no. 772195, the Deutsche Forschungsgemeinschaft (DFG, German Research Foundation) within the Germany's Excellence Strategy under grant no. EXC-2111-390814868, and the Bavarian Hightech Agenda within the Munich Quantum Valley project EQAP. Open access funding enabled and organized by Projekt DEAL.

**Conflict of interest**

The authors declare no conflict of interest.

**Author contributions**

N.F.Z. fabricated and measured all samples and analyzed the data. S.S. fabricated selected samples, contributed to characterization and prepared chemically functionalized SWCNT dispersions. S.L. prepared (7,5) SWCNT dispersions and performed Raman mapping of *h*-BN flakes. A.A.E.Y. contributed to temperature-dependent PL measurements and measurements on *h*-BN flakes. T.W. exfoliated h-BN crystals and S.Z. performed hyperspectral PL mapping of *h*-BN flakes under the supervision of A.H.. H.L. and B.S.F. prepared ATPE-sorted (6,5) SWCNT dispersions. J.Z. conceived and supervised the project. N.F.Z. and J.Z. wrote the manuscript with input from all authors.

# Supporting Information

## Near-intrinsic photo- and electroluminescence from single-walled carbon nanotube thin films on BCB-passivated surfaces

*Nicolas F. Zorn, Simon Settele, Shen Zhao, Sebastian Lindenthal, Abdurrahman Ali El Yumin, Tim Wedl, Han Li, Benjamin S. Flavel, Alexander Högele, and Jana Zaumseil**

## Table of Contents

## Absorption Spectrum of (6,5) SWCNT Dispersion

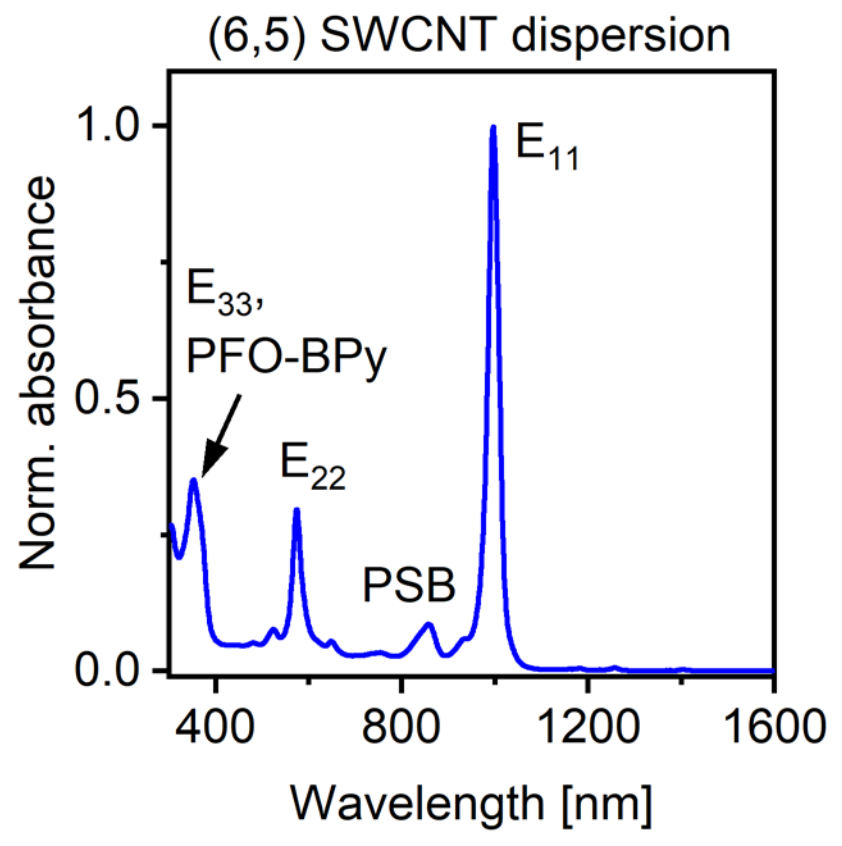

**Figure S1.** Normalized absorption spectrum of a toluene dispersion of (6,5) SWCNTs wrapped with the polyfluorene copolymer PFO-BPy (after removal of excess polymer). The main excitonic transitions ($E_{11}$, $E_{22}$ and $E_{33}$), the $E_{11}$ phonon sideband (PSB) and the absorption of the residual wrapping polymer are labelled.

**PL Spectra of (6,5) SWCNT Thin Films on Glass and Si/$SiO_2$ Substrates: Effect of Annealing Temperature**

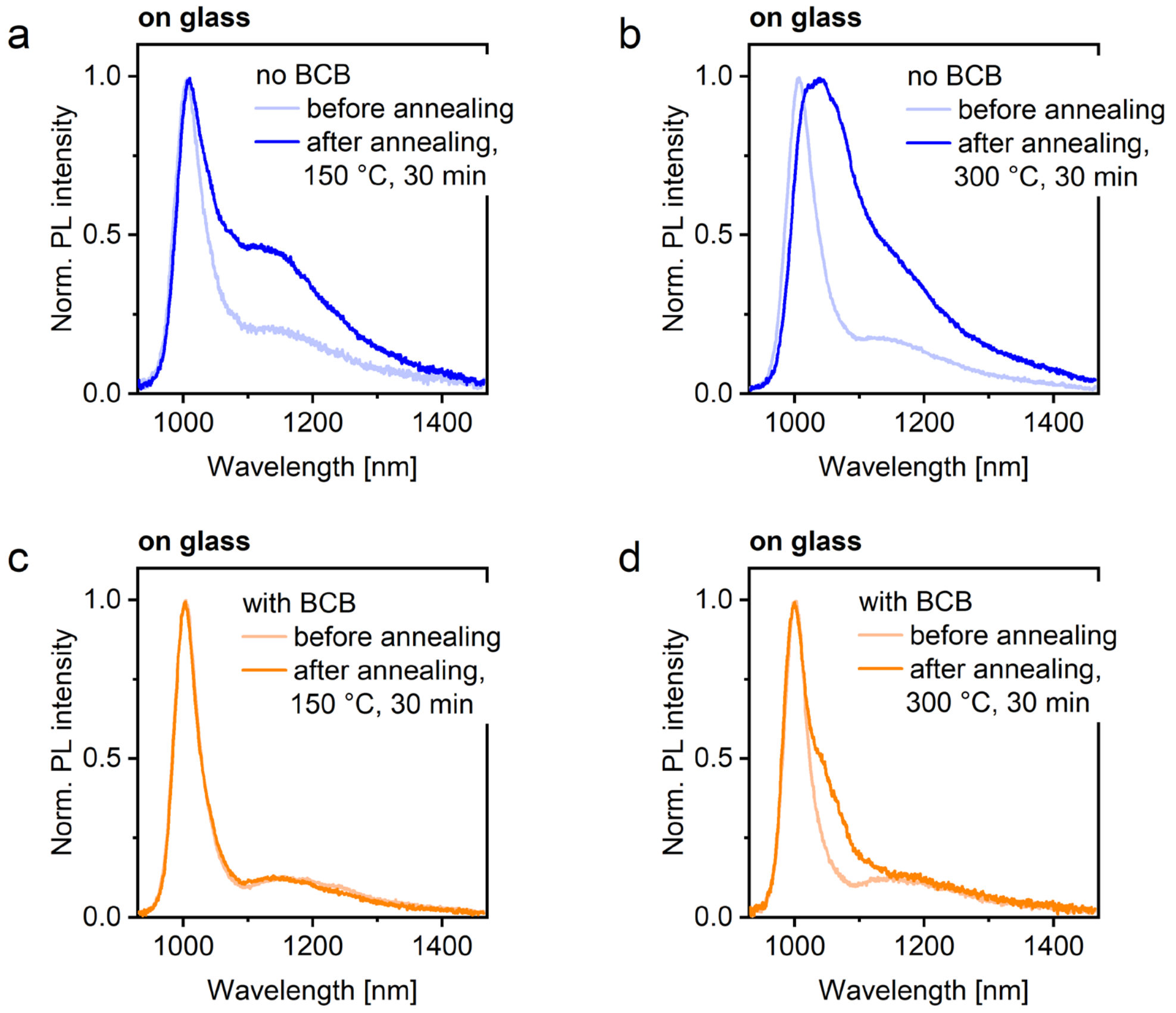

**Figure S2. (a,b)** Normalized PL spectra of (6,5) SWCNT thin films on glass substrates without BCB layer before and after annealing in inert atmosphere at **(a)** 150 °C and **(b)** 300 °C for 30 min. **(c,d)** Normalized PL spectra of (6,5) SWCNT thin films on glass substrates with BCB layer (~75 nm) before and after annealing in inert atmosphere at **(c)** 150 °C and **(d)** 300 °C for 30 min.

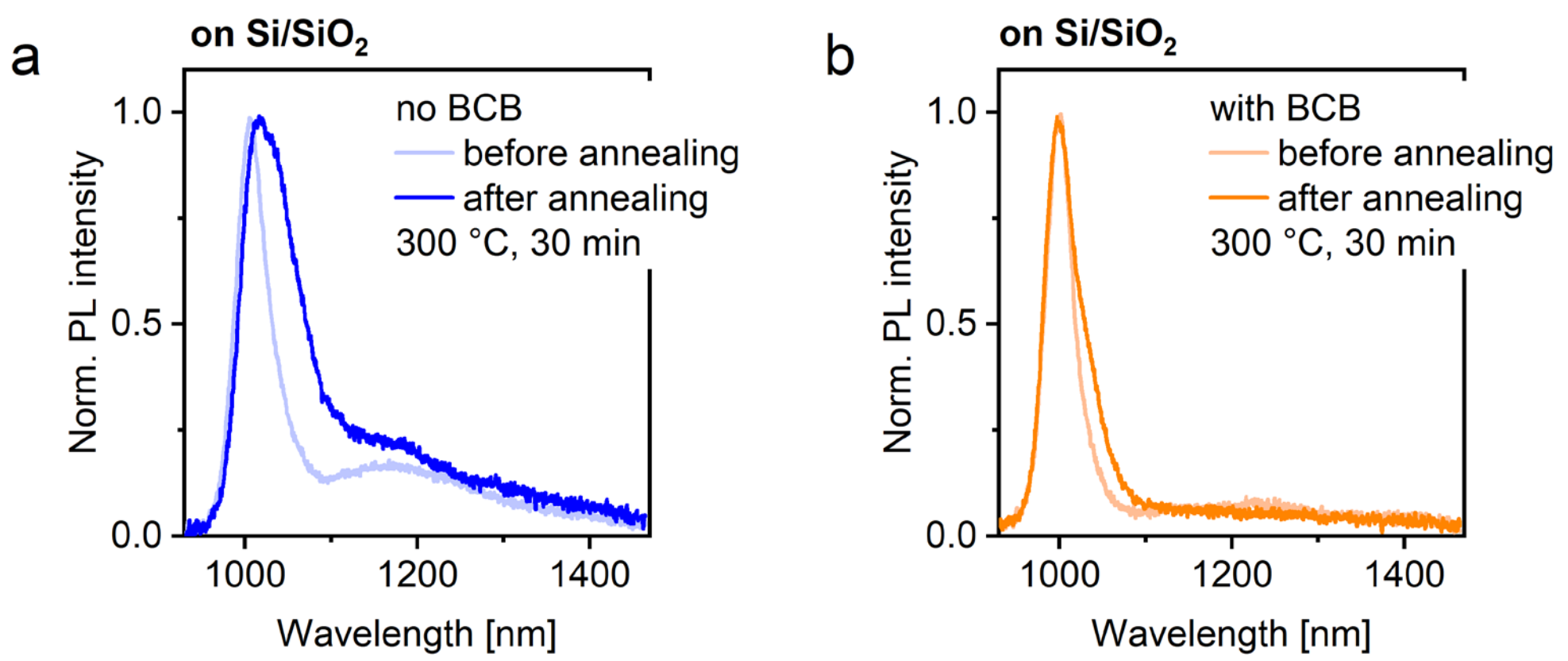

**Figure S3.** Normalized PL spectra of (6,5) SWCNT thin films before and after annealing in inert atmosphere at 300 °C for 30 min, **(a)** on $Si/SiO_2$ substrates without BCB layer and **(b)** on $Si/SiO_2$ substrates with BCB layer (~65 nm).

## Absorption Spectrum of BCB Precursor Solution

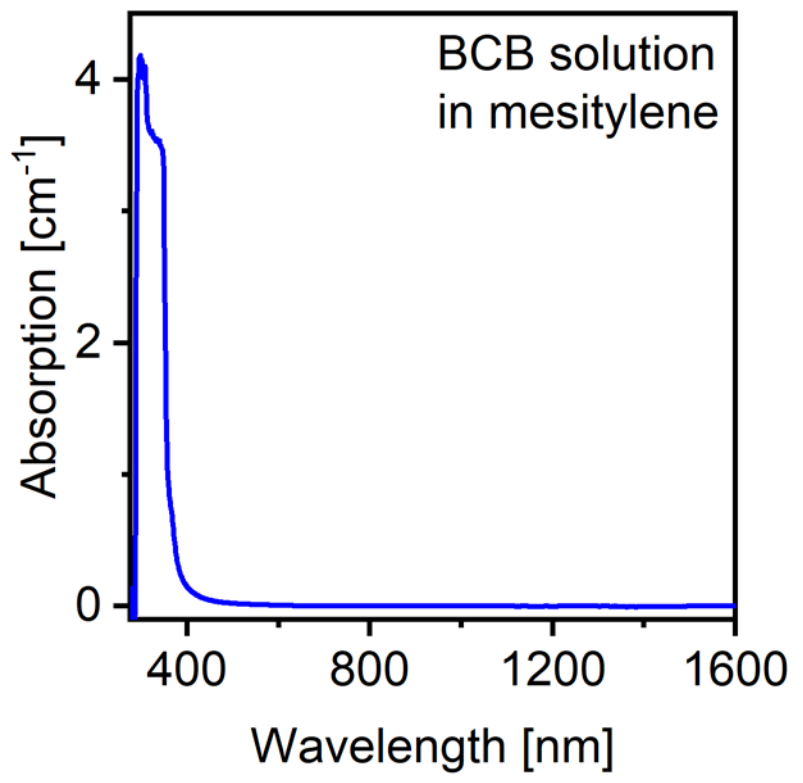

**Figure S4.** Absorption spectrum of a diluted solution of BCB precursor in mesitylene. No significant absorption above 500 nm is visible in the spectrum.

**PL Spectra of (6,5) SWCNT Thin Films on Glass Substrates: Effect of BCB Thickness and SWCNT Network Density**

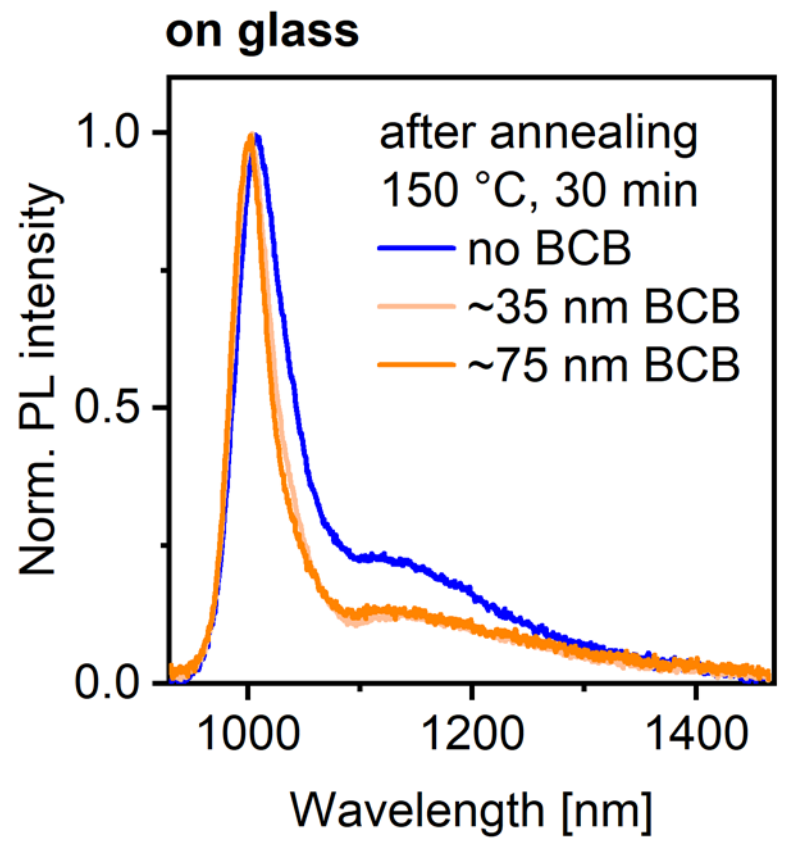

**Figure S5.** Normalized PL spectra of (6,5) SWCNT thin films on glass substrates without BCB layer, with ~35 nm BCB layer, and with ~75 nm BCB layer after annealing in inert atmosphere at 150 °C for 30 min.

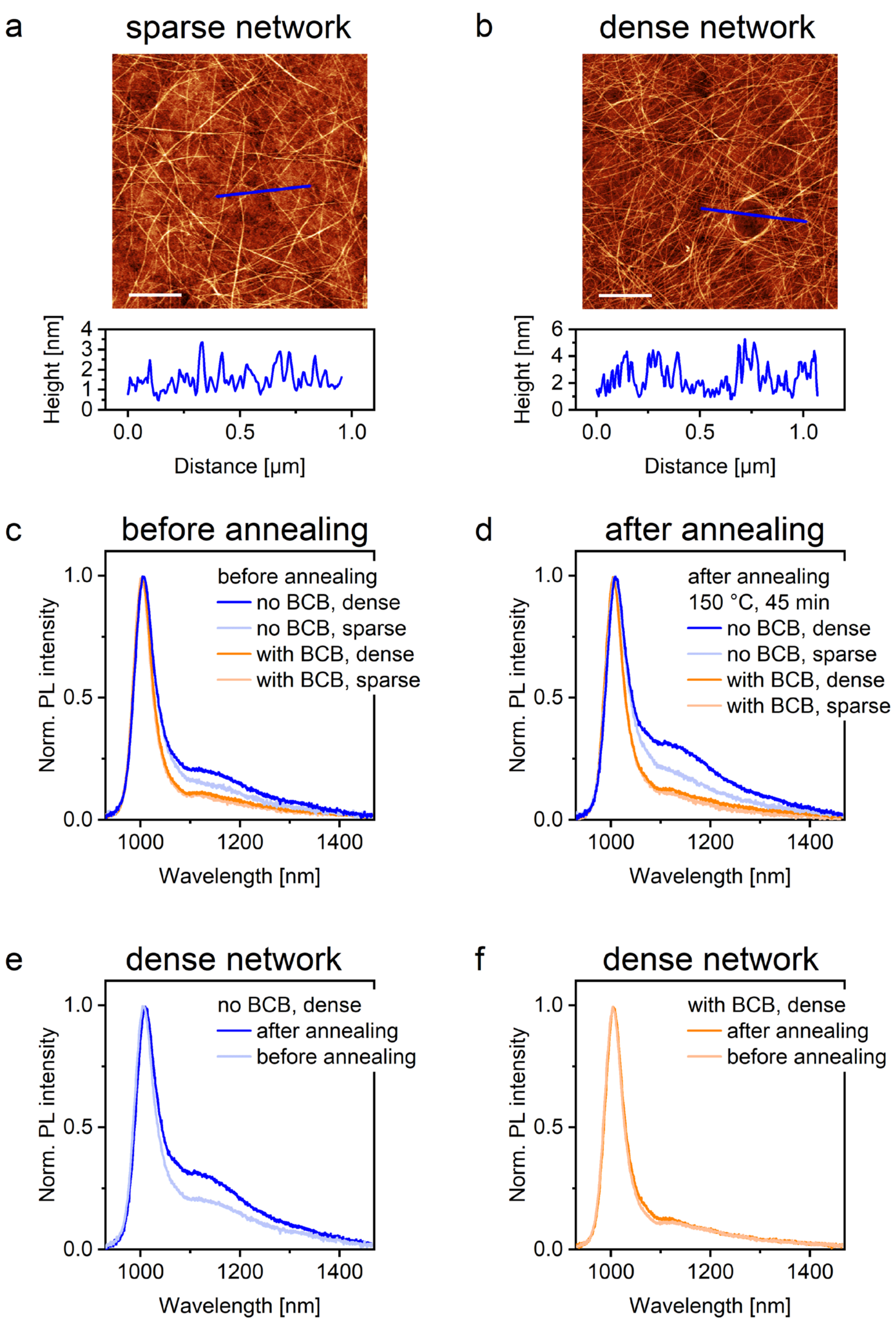

**Figure S6.** Effect of SWCNT network thickness. **(a)** Atomic force micrograph of a sparse network and **(b)** of a dense network of (6,5) SWCNTs on glass substrates and height profiles along the lines indicated in blue color. Scale bars are 500 nm. **(c,d)** Normalized PL spectra of sparse and dense SWCNT networks on glass substrates without and with ~75 nm BCB layer **(c)** before annealing and **(d)** after annealing in inert atmosphere at 150 °C for 45 min. **(e,f)** Normalized PL spectra of dense SWCNT networks on glass substrates before and after annealing in inert atmosphere at 150 °C for 45 min on glass substrates **(e)** without BCB layer and **(f)** with ~75 nm BCB layer.

**Raman Spectra of (6,5) SWCNT Thin Films on Glass and Si/SiO$_2$ Substrates**

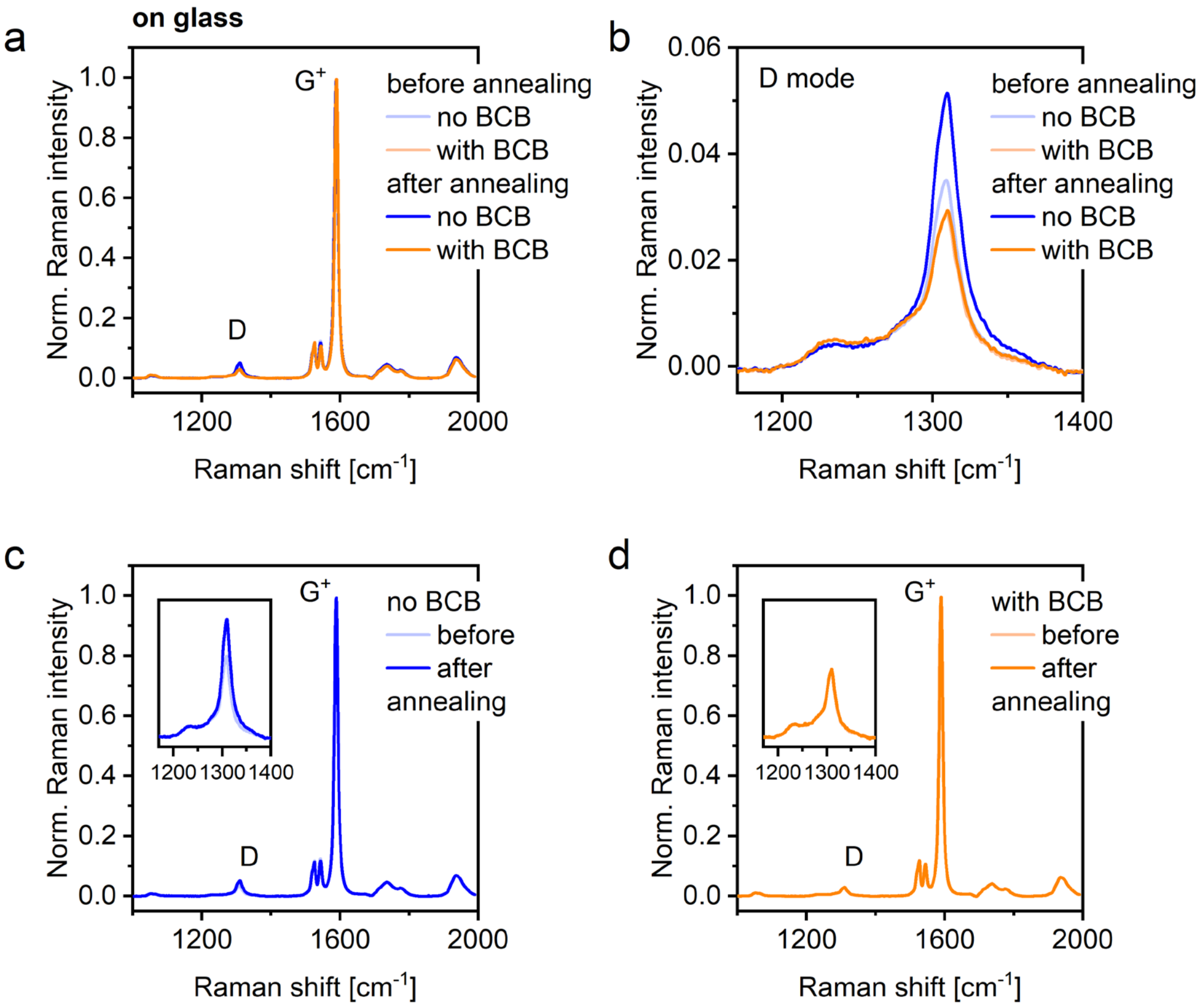

**Figure S7.** **(a)** Normalized Raman spectra (532 nm excitation) of (6,5) SWCNT thin films on glass substrates without and with BCB layer (~75 nm) before and after annealing in inert atmosphere at 150 °C for 30 min. Spectra are averaged from ~2600 individual measurements obtained by mapping an area of 100 x 100 µm$^2$. **(b)** Zoom-in on the defect-related D mode. **(c,d)** Raman spectra of (6,5) SWCNT thin films on glass substrates **(c)** without and **(d)** with BCB layer. The insets show a zoom-in on the D mode peak.

**Table S1.** Integrated Raman D/G$^+$ ratios of (6,5) SWCNT thin films shown in **Figure S7**.

| Sample | Raman D/G$^+$ ratio before annealing | Raman D/G$^+$ ratio after annealing (150 °C, 30 min) |
|---|---|---|
| No BCB | 0.079 | 0.109 |
| With BCB | 0.077 | 0.079 |

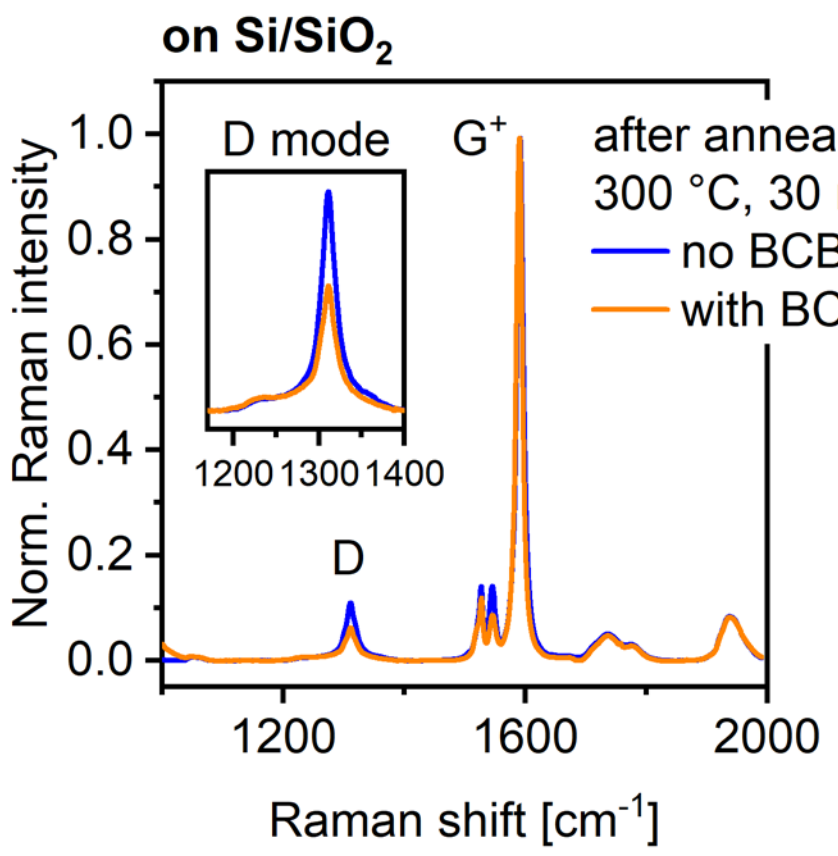

**Figure S8.** Normalized Raman spectra (532 nm excitation) of (6,5) SWCNT thin films on $Si/SiO_2$ without and with BCB layer (~65 nm) after annealing in inert atmosphere at 300 °C for 30 min. Note that the minor discrepancies in the relative intensities of the $G^-$ (~1526 $cm^{-1}$) and $E_2$ (~1546 $cm^{-1}$) Raman modes with respect to the $G^+$ mode (~1590 $cm^{-1}$) might be due to unintentional p-doping of SWCNT films in air.[1]

**Table S2.** Integrated Raman $D/G^+$ ratios of (6,5) SWCNT thin films shown in **Figure S8**.

| Sample | Raman $D/G^+$ ratio after annealing (300 °C, 30 min) |
|---|---|
| No BCB | 0.20 |
| With BCB | 0.14 |

## Temperature-Dependent PL Spectra of (6,5) SWCNT Thin Films on Glass Substrates

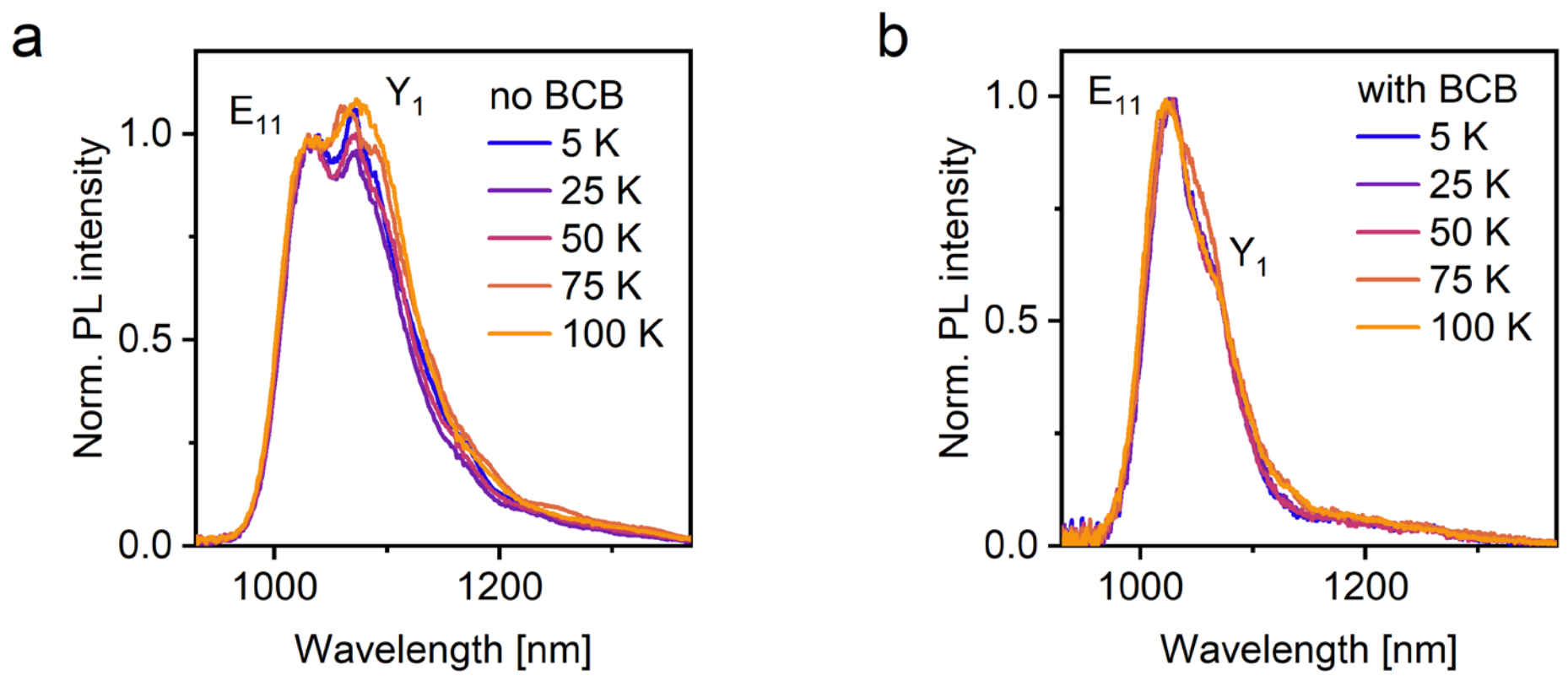

**Figure S9.** Normalized temperature-dependent PL spectra of (6,5) SWCNT thin films on glass substrates **(a)** without and **(b)** with ~80 nm BCB layer between 5 K and 100 K. Samples were annealed in inert atmosphere at 150 °C for 45 min.

## Excitation Power-Dependent PL Spectra of (6,5) SWCNT Thin Films on Glass Substrates

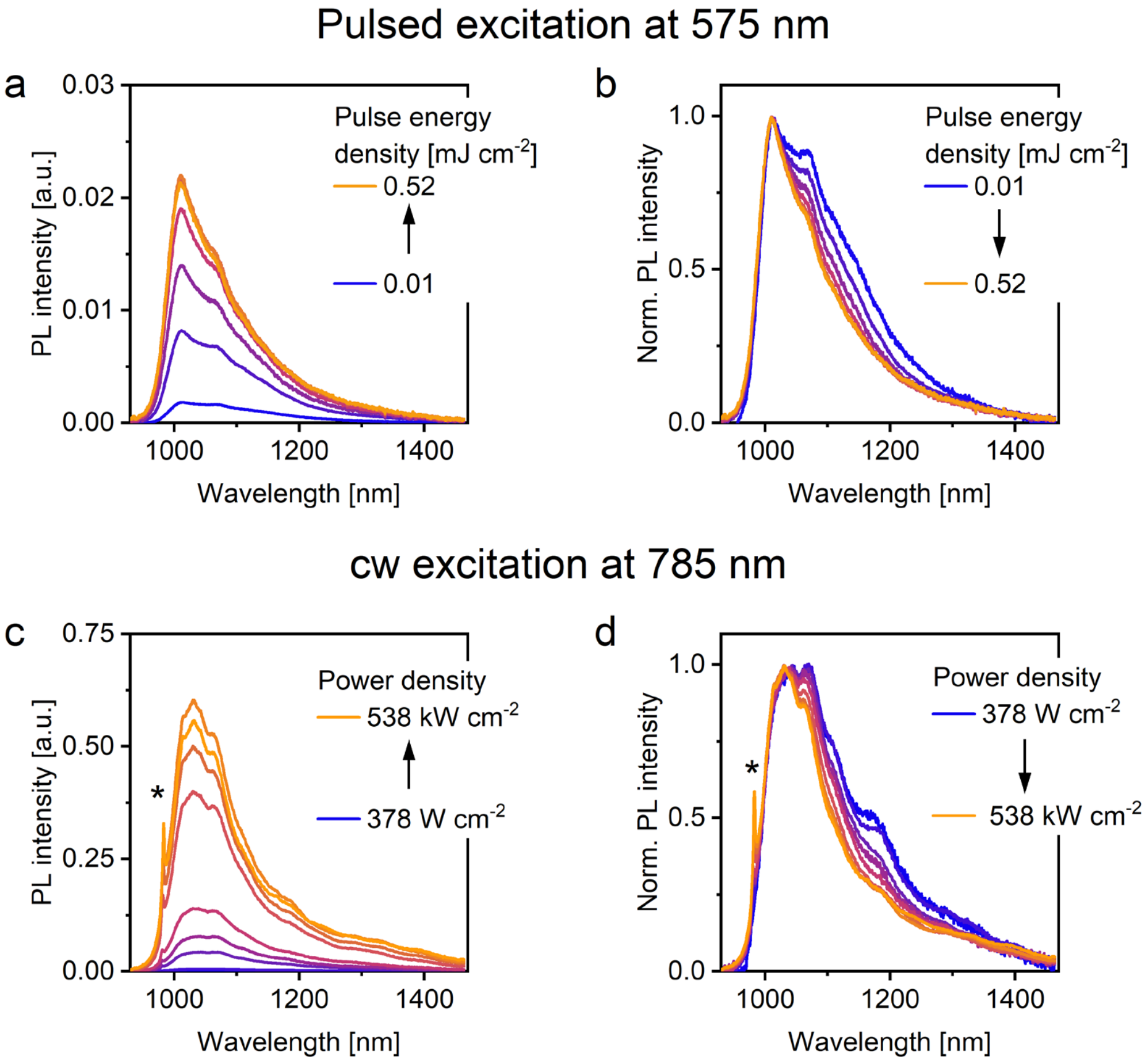

**Figure S10.** Excitation power-dependence of PL sideband emission. **(a)** Absolute and **(b)** normalized PL spectra under pulsed excitation at 575 nm of a (6,5) SWCNT thin film on glass without BCB after annealing in inert atmosphere at 300 °C for 30 min. **(c)** Absolute and **(d)** normalized PL spectra of the same sample under continuous wave (cw) excitation at 785 nm. The asterisk marks the Raman 2D band of (6,5) SWCNTs.

## PL Spectra of (7,5) SWCNT Thin Films on Glass and Si/SiO$_2$ Substrates

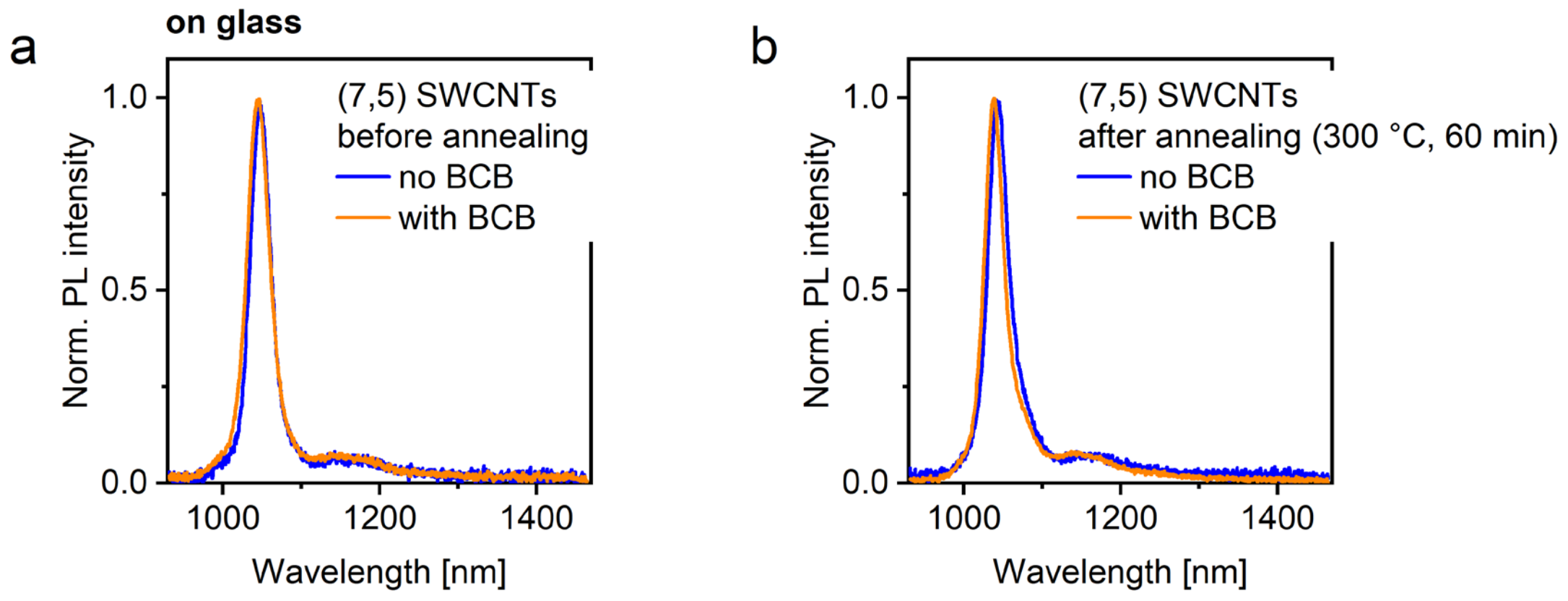

**Figure S11.** Normalized PL spectra of (7,5) SWCNT thin films on glass substrates without and with ~75 nm BCB layer **(a)** before annealing and **(b)** after annealing in inert atmosphere at 300 °C for 60 min.

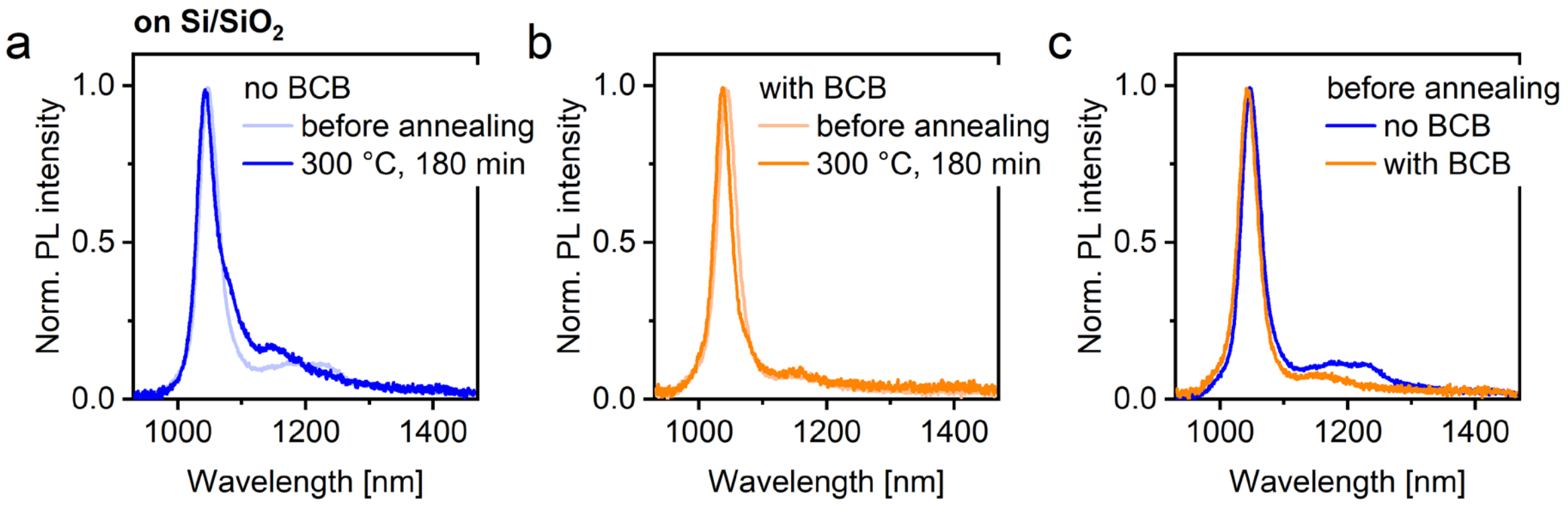

**Figure S12.** Normalized PL spectra of (7,5) SWCNT thin films on Si/SiO$_2$ substrates **(a)** without BCB layer and **(b)** with ~75 nm BCB layer before and after annealing in inert atmosphere at 300 °C for 180 min. **(c)** Normalized PL spectra without and with BCB layer before annealing.

## PL and Raman Spectra of (6,5) SWCNTs on $Si/SiO_2$ Substrates with *h*-BN flakes

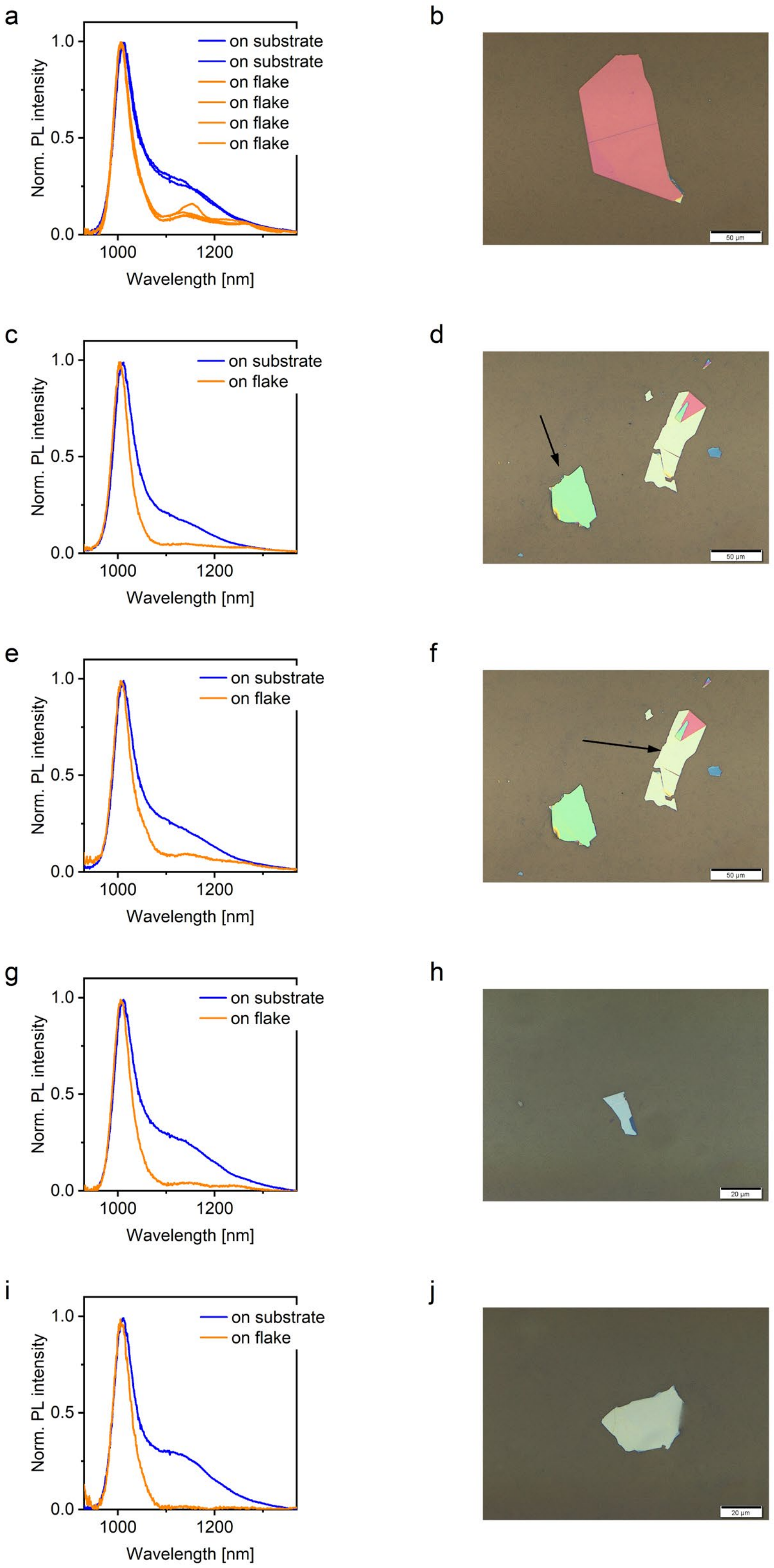

**Figure S13.** PL spectroscopy of a (6,5) SWCNT thin film on a $Si/SiO_2$ substrate with *h*-BN flakes. **(a,c,e,g,i)** Normalized PL spectra on the substrate (blue lines) and on *h*-BN flakes (orange lines). **(b,d,f,h,j)** Optical micrographs of the corresponding *h*-BN flakes.

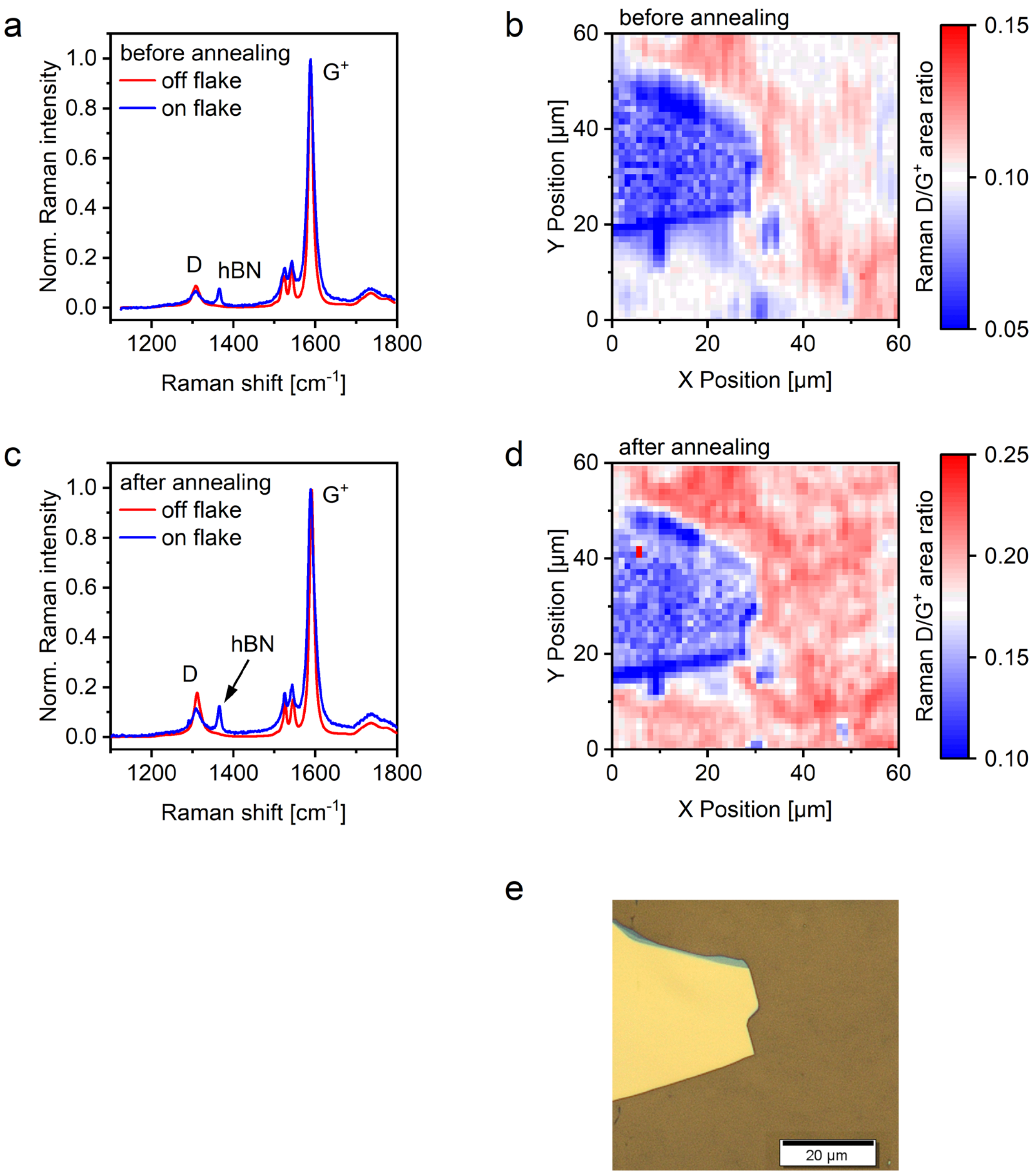

**Figure S14.** Raman spectroscopy (532 nm excitation) of a (6,5) SWCNT thin film on a Si/$SiO_2$ substrate with a ~80 nm *h*-BN flake. **(a)** Normalized, averaged Raman spectra on the flake (blue line) and on the substrate (red line) before annealing and **(b)** map of the integrated D-to-$G^+$ peak area ratio. **(c)** Normalized, averaged Raman spectra on the same flake (blue line) and on the substrate (red line) after annealing in inert atmosphere at 300 °C for 30 min and **(d)** map of the integrated D-to-$G^+$ peak area ratio. Please note the different color scales in (b) and (d). **(e)** Optical microscope image of the corresponding *h*-BN flake. The scale bar is 20 µm.

## (6,5) SWCNT Network FETs on Glass Substrates without and with BCB

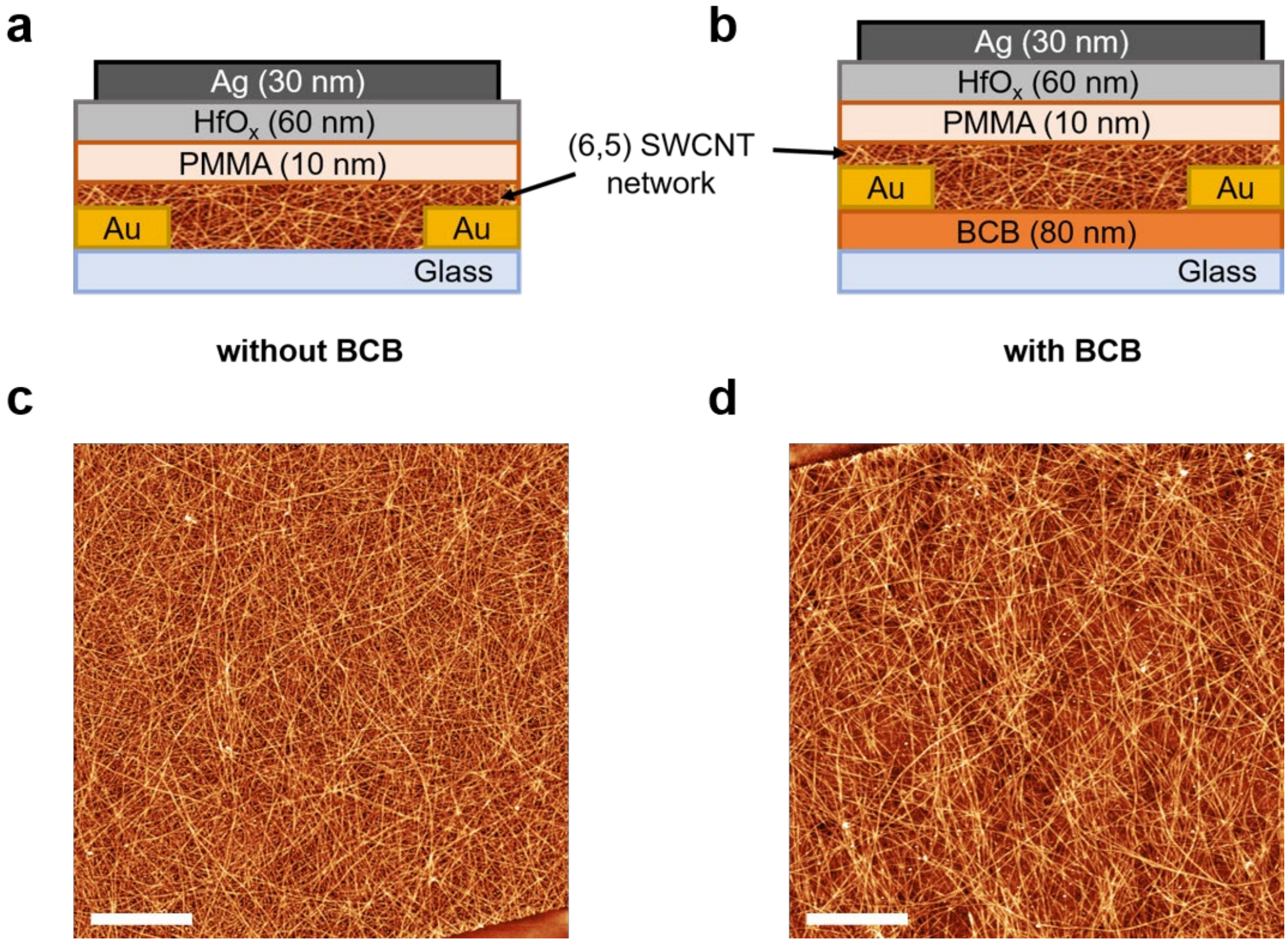

**Figure S15.** Schematic layout of (6,5) SWCNT network field-effect transistors **(a)** without BCB layer and **(b)** with ~80 nm BCB layer passivating the glass substrate. Layer thicknesses are not to scale. **(c,d)** Representative atomic force micrographs of SWCNT networks **(c)** on glass and **(d)** on the BCB layer. Scale bars are 1 µm.

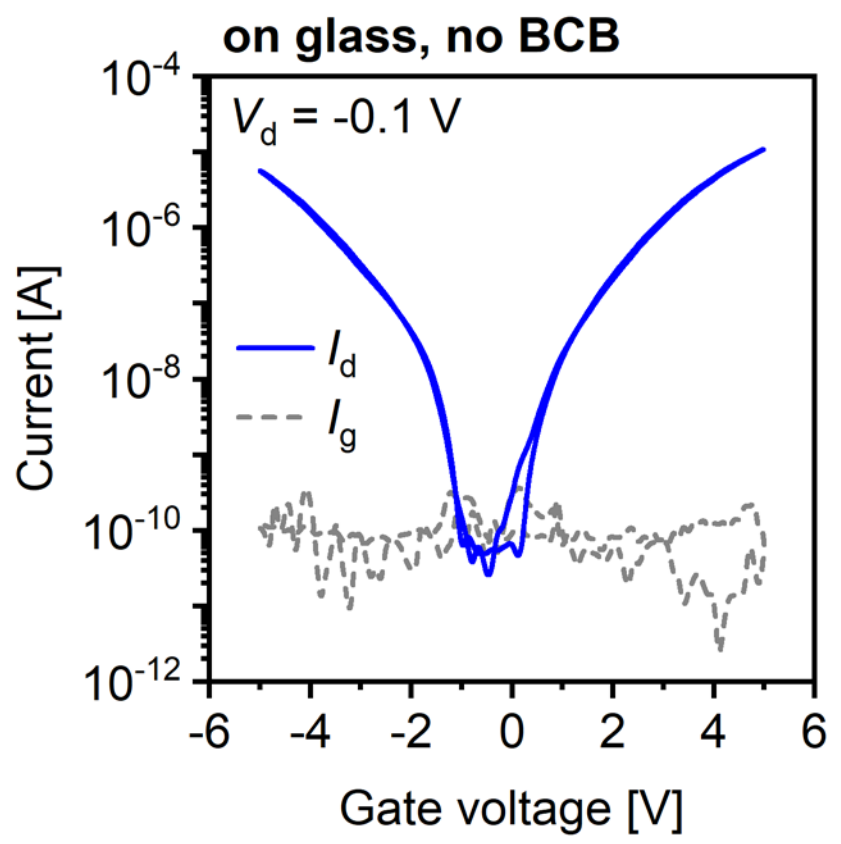

**Figure S16.** Transfer characteristics of a (6,5) SWCNT network FET on glass without BCB in the linear regime ($V_d$ = -0.1 V). The blue solid line is the drain current $I_d$, the gray dashed line is the gate leakage current $I_g$.

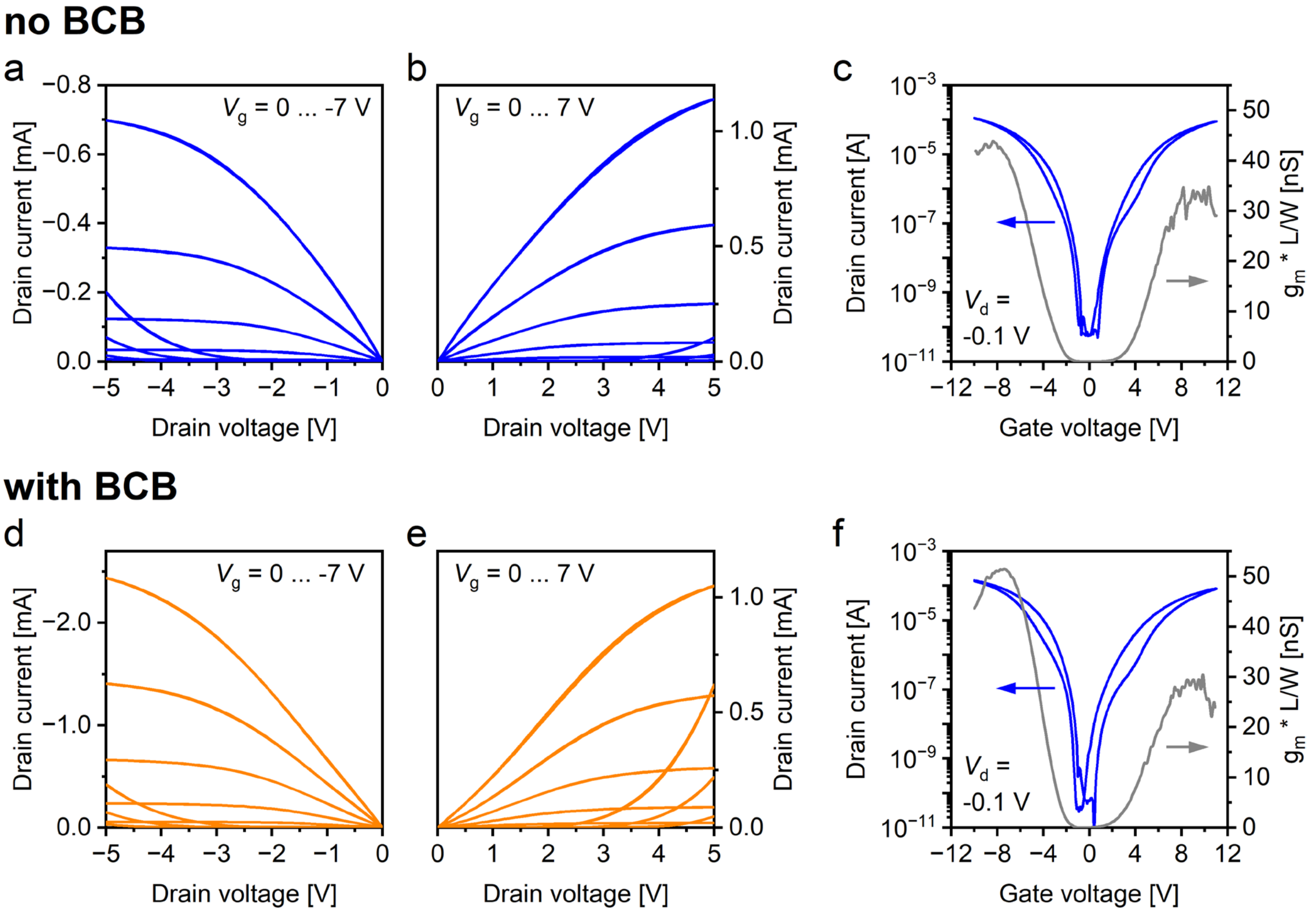

**Figure S17.** Output characteristics of (6,5) SWCNT network FETs on glass **(a,b)** without BCB and **(d,e)** with ~80 nm BCB. **(c,f)** Linear transfer characteristics (blue lines; $V_d$ = -0.1 V) and gate voltage-dependent transconductance normalized to $L/W$ (gray lines).

**Table S3.** Average linear field-effect mobilities and peak transconductance values $g_m$ (normalized to $L/W$) for holes and electrons of (6,5) SWCNT network FETs on glass substrates without and with ~80 nm BCB layer. Samples were annealed in inert atmosphere at 150 °C for 45 min.

| Sample | Avg. hole mobility | Avg. electron mobility | Norm. transconductance $g_m$ * L/W | |
|---|---|---|---|---|
| | [cm$^2$(Vs)$^{-1}$] | [cm$^2$(Vs)$^{-1}$] | for holes [nS] | for electrons [nS] |
| No BCB | 3.45 ± 0.17 | 2.43 ± 0.13 | 45.4 | 32.0 |
| With BCB | 3.85 ± 0.26 | 2.14 ± 0.30 | 50.1 | 27.8 |

## Electroluminescence (EL) from (6,5) SWCNT Network FETs

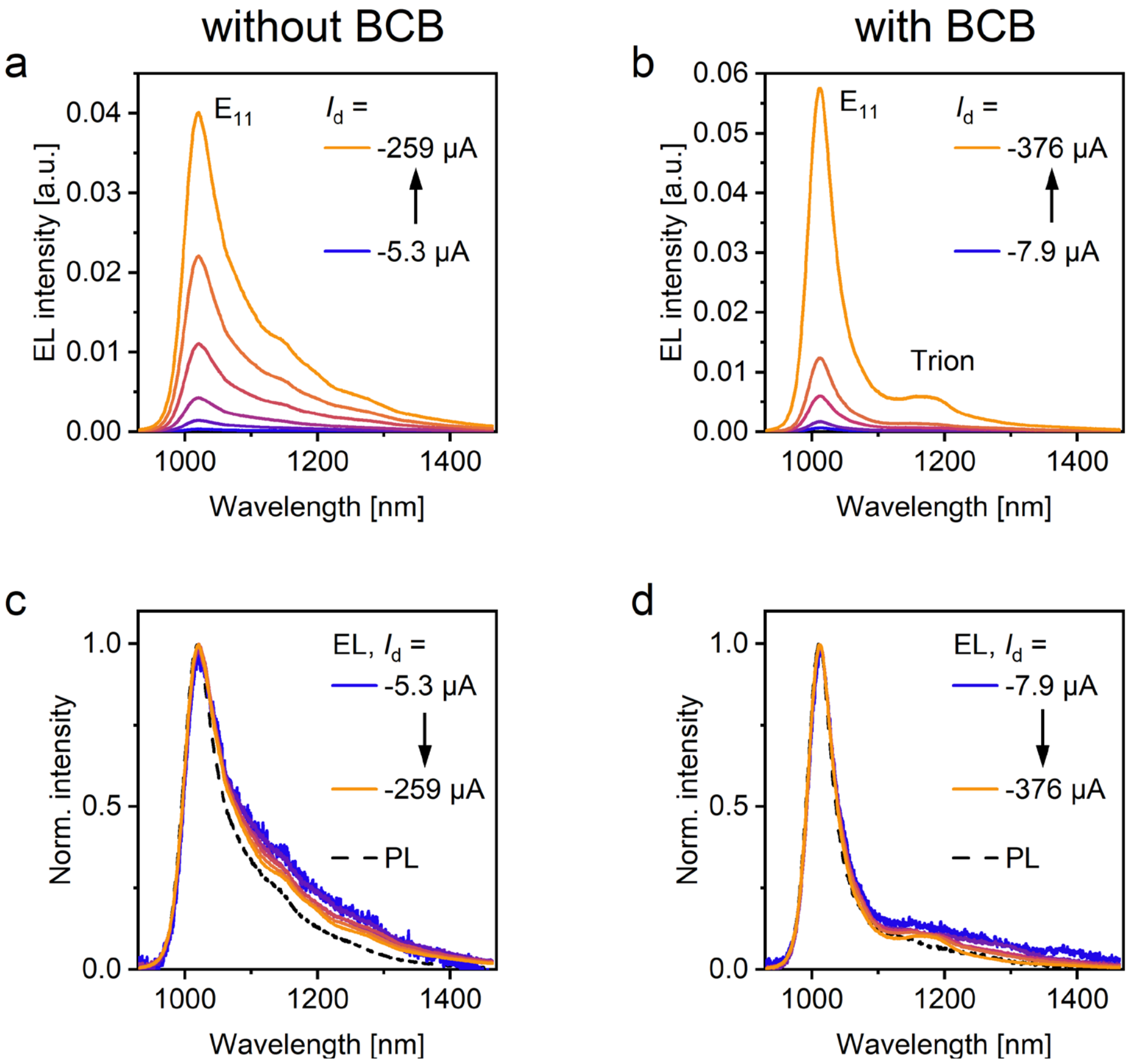

**Figure S18. (a,b)** Absolute and **(c,d)** normalized EL spectra of (6,5) SWCNT network FETs on glass substrates for different drain currents $I_d$. Data is shown for samples **(a,c)** without and **(b,d)** with ~80 nm BCB layer. Samples were annealed in inert atmosphere at 150 °C for 45 min before deposition of PMMA/$HfO_x$. In **(c,d)**, dashed black lines represent normalized PL spectra from the same spot.

## Gate Voltage-Dependent PL from (6,5) SWCNT Network FETs

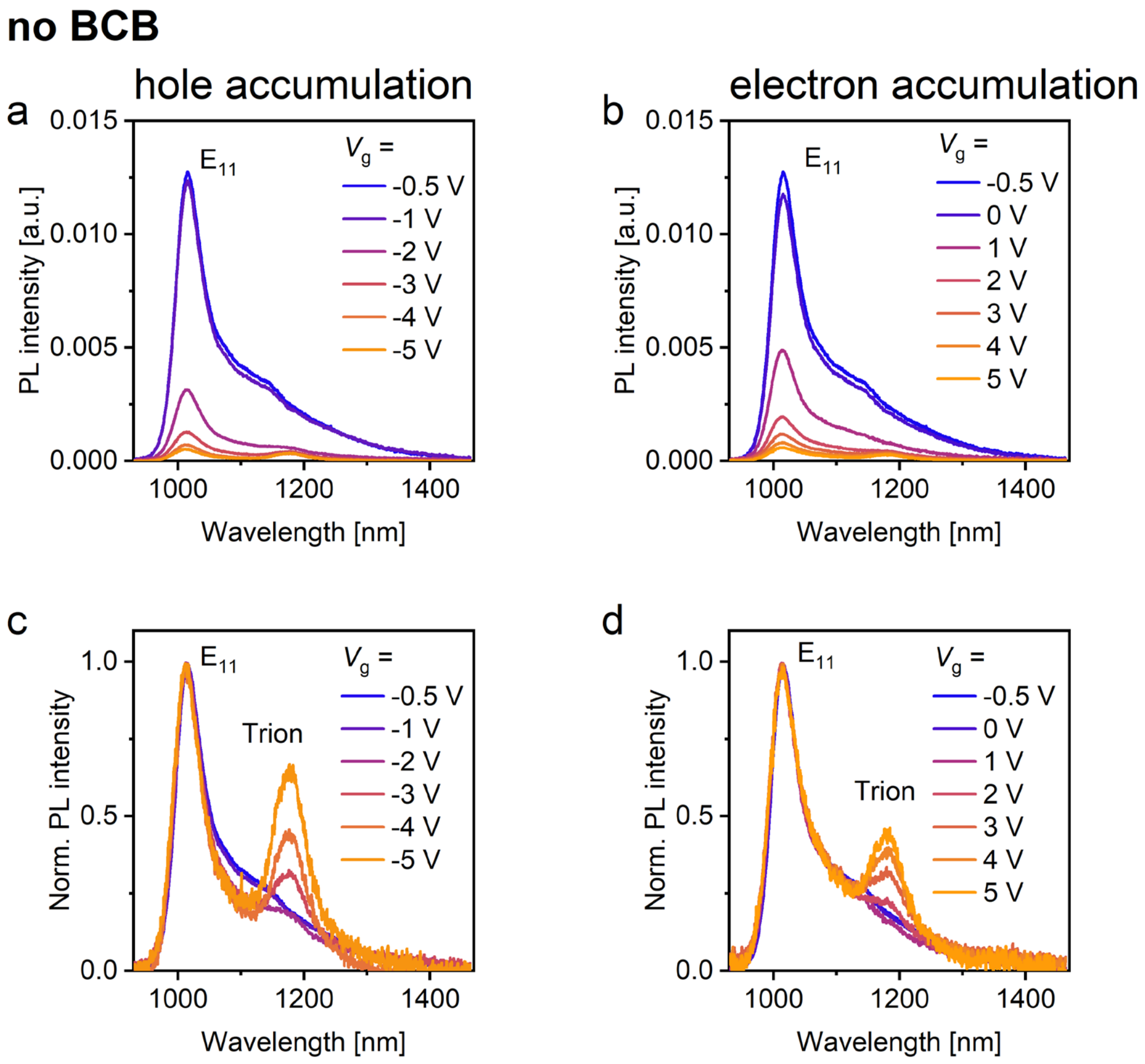

**Figure S19. (a,b)** Absolute and **(c,d)** normalized PL spectra of (6,5) SWCNT network FETs on glass substrates without BCB layer for different gate voltages $V_g$. Spectra were acquired in **(a,c)** hole and **(b,d)** electron accumulation. Samples were annealed in inert atmosphere at 150 °C for 45 min before deposition of PMMA/$HfO_x$.

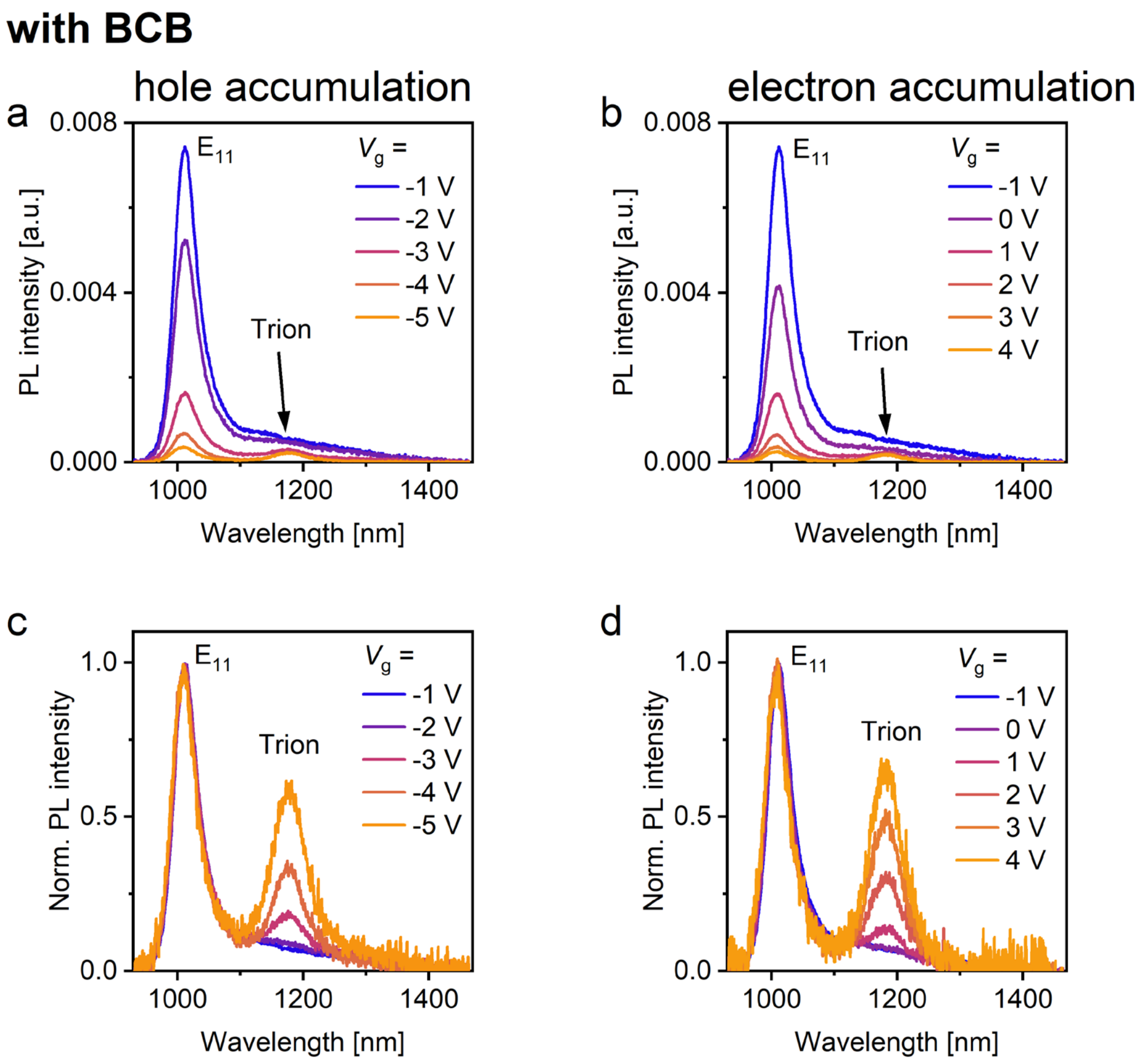

**Figure S20. (a,b)** Absolute and **(c,d)** normalized PL spectra of (6,5) SWCNT network FETs on glass substrates with ~80 nm BCB layer for different gate voltages $V_g$. Spectra were acquired in **(a,c)** hole and **(b,d)** electron accumulation. Samples were annealed in inert atmosphere at 150 °C for 45 min before deposition of PMMA/$HfO_x$.

## $sp^3$-Functionalized (6,5) SWCNT Thin Films on Glass Substrates

**Experimental Details:** Covalent $sp^3$ functionalization of polymer-wrapped (6,5) SWCNTs in organic solvents was performed according to recently published protocols.[2,3] Reactions were carried out at a (6,5) SWCNT concentration of 0.54 µg mL$^{-1}$, corresponding to an optical density of 0.3 cm$^{-1}$ at the $E_{11}$ absorption peak. $E_{11}$* defects were introduced through reaction with 4-nitrobenzenediazonium tetrafluoroborate (1.1 mmol L$^{-1}$) in a solvent mixture of toluene and acetonitrile (80:20 vol-%) with 18-crown-6 (7.6 mmol L$^{-1}$) as phase-transfer agent. After ~16 h in the dark at room temperature, the reaction was stopped by filtration of the dispersion through a PTFE membrane filter (Merck Millipore JVWP, pore size 0.1 µm). The filter cake was thoroughly washed with acetonitrile and toluene to remove unreacted diazonium salt. Further red-shifted $E_{11}{}^{*-}$ defects were created through reaction with 2-iodoaniline (29.3 mmol L$^{-1}$) and potassium *tert*-butoxide (KO$^t$Bu, 58.6 mmol L$^{-1}$) in the presence of dimethyl sulfoxide (DMSO, anhydrous) and tetrahydrofuran (THF, anhydrous). The volumetric fractions were 83.3:8.3:8.3 vol-% toluene:DMSO:THF. The reaction was allowed to proceed in the dark under vigorous stirring for 60 min before being stopped by vacuum filtration over a PTFE membrane filter (Merck Millipore JVWP, pore size 0.1 µm) and washing with methanol and toluene to remove unreacted compounds and by-products. Bath sonication of the filter cakes for 30 min in fresh toluene with 1,10-phenantroline (2.8 mmol L$^{-1}$) as stabilizer[4] yielded dispersions of $sp^3$-functionalized (6,5) SWCNTs, which were immediately used for characterization or thin film deposition as described in the Methods section of the main text.

PL spectra of (6,5) SWCNT thin films with $sp^3$ defects on glass substrates under pulsed excitation at 575 nm were acquired as detailed in the Methods section of the main manuscript. On the same setup, time-resolved PL measurements were conducted in a time-correlated single-photon counting (TCSPC) scheme. The emission was spectrally selected with a spectrograph (Princeton Instruments Acton SpectraPro SP2358, grating 150 lines mm$^{-1}$) and focused onto an InGaAs/InP avalanche photodiode (Micro Photon Devices) with a NIR-optimized 20× objective (N.A. 0.40, Mitutoyo). TCSPC histograms were calculated using a counting module (PicoQuant PicoHarp 300) and fitted with biexponential functions in a reconvolution procedure. The fast, instrument-limited $E_{11}$ PL decay of a thick, drop-cast (6,5) SWCNT film served as instrument response function.

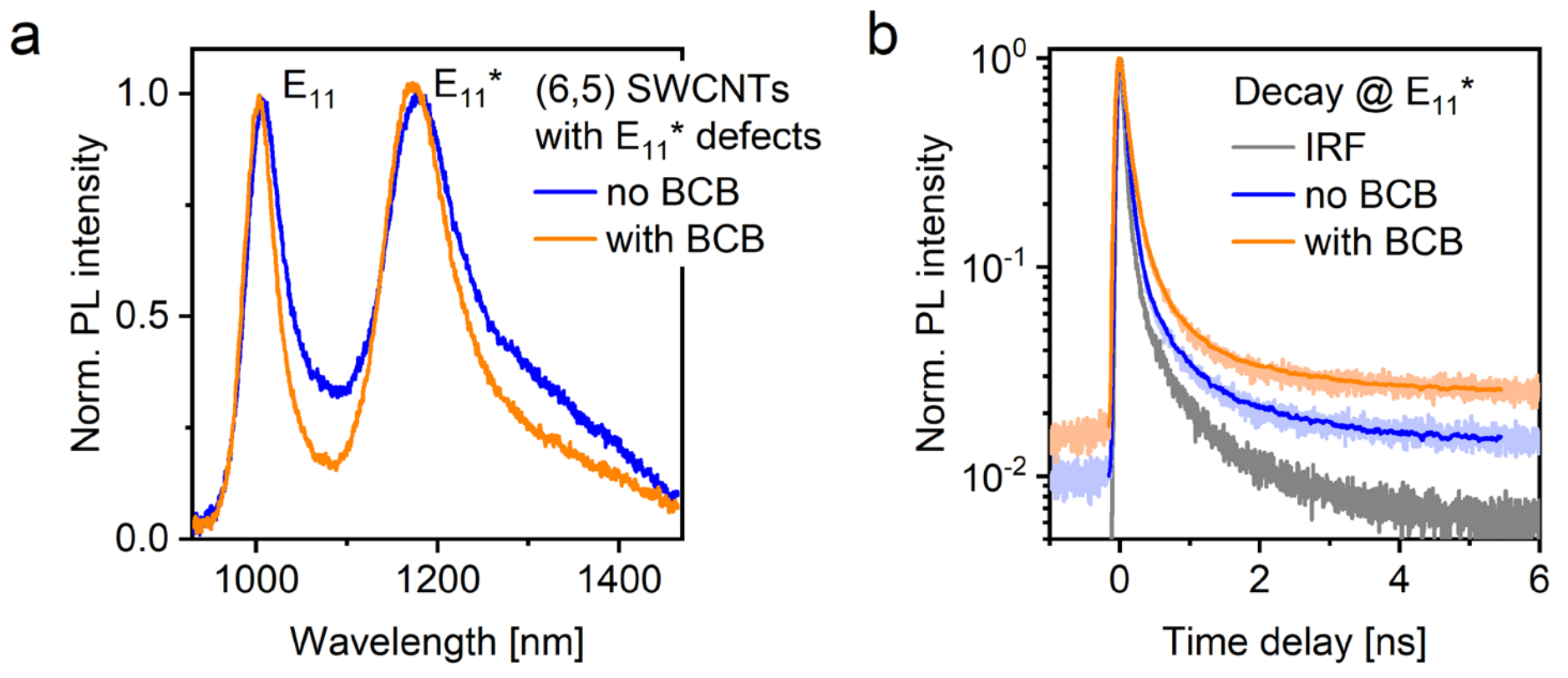

**Figure S21. (a)** Normalized PL spectra of (6,5) SWCNT thin films with $E_{11}$* *sp*$^3$ defects on glass substrates without and with ~75 nm BCB layer. Samples were annealed in inert atmosphere at 150 °C for 45 min. **(b)** PL decay at the $E_{11}$* emission wavelength (1170 nm) for thin films on glass and on BCB. The instrument response function (IRF) is shown in gray.

**Table S4.** Long ($\tau_{long}$) and short ($\tau_{short}$) lifetime components, normalized amplitudes ($A_{long}$, $A_{short}$), and amplitude-averaged lifetimes ($\tau_{ampl.-avg.}$) obtained by fitting the PL decay of (6,5) SWCNT dispersions and thin films with $E_{11}$* *sp*$^3$ defects at the $E_{11}$* emission wavelength (1170 nm) with a biexponential function.

| Sample | $\tau_{long}$ [ps] | $\tau_{short}$ [ps] | $A_{long}$ | $A_{short}$ | $\tau_{ampl.-avg.}$ [ps] |
|---|---|---|---|---|---|
| Dispersion | 433 | 156 | 0.24 | 0.76 | 222 |
| Thin film, no BCB | 56 | 26 | 0.21 | 0.79 | 32 |
| Thin film, with BCB | 131 | 48 | 0.14 | 0.86 | 60 |

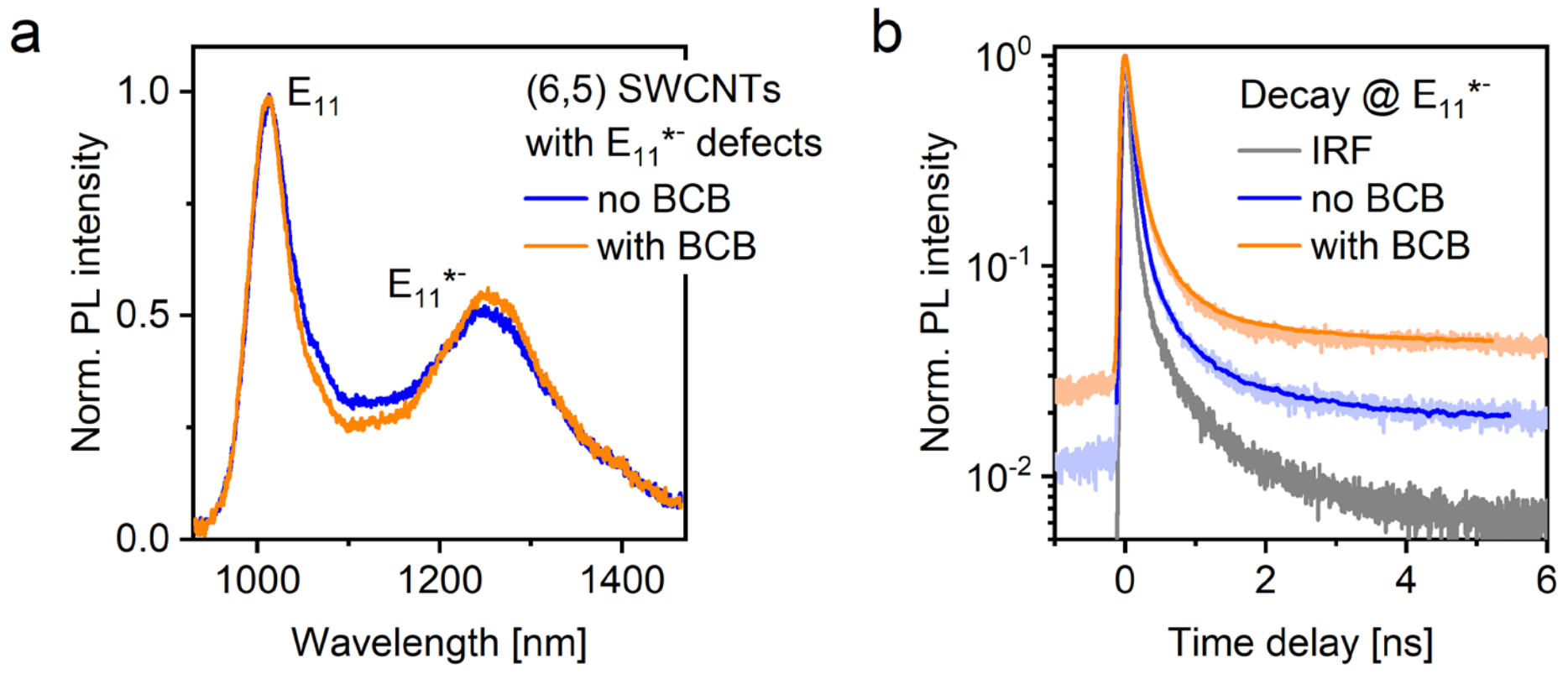

**Figure S22. (a)** Normalized PL spectra of (6,5) SWCNT thin films with $E_{11}$*- *sp*$^3$ defects on glass substrates without and with ~75 nm BCB layer. Samples were annealed in inert atmosphere at 150 °C for 45 min. **(b)** PL decay at the $E_{11}$*- emission wavelength (1250 nm) for thin films on glass and on BCB. The instrument response function (IRF) is shown in gray.

**Table S5.** Long ($\tau_{long}$) and short ($\tau_{short}$) lifetime components, normalized amplitudes ($A_{long}$, $A_{short}$), and amplitude-averaged lifetimes ($\tau_{ampl.-avg.}$) obtained by fitting the PL decay of (6,5) SWCNT dispersions and thin films with $E_{11}$*- *sp*$^3$ defects at the $E_{11}$*- emission wavelength (1250 nm) with a biexponential function.

| Sample | $\tau_{long}$ [ps] | $\tau_{short}$ [ps] | $A_{long}$ | $A_{short}$ | $\tau_{ampl.-avg.}$ [ps] |
|---|---|---|---|---|---|
| Dispersion | 562 | 173 | 0.47 | 0.53 | 355 |
| Thin film, no BCB | 89 | 33 | 0.12 | 0.88 | 40 |
| Thin film, with BCB | 189 | 55 | 0.09 | 0.91 | 67 |